\documentclass[11pt,onecolumn]{article}

\usepackage{bm}
\usepackage{amsfonts}
\usepackage{amsmath}
\usepackage{amssymb}
\usepackage{cite}
\usepackage[amsmath,thmmarks]{ntheorem}

\usepackage{color, soul}
\usepackage{epic, eepic}
\usepackage[top=1in, bottom=1in, left=1.25in, right=1.25in]{geometry}
\usepackage{appendix}

\usepackage[dvips]{graphicx}
\usepackage{color, soul}
\usepackage{array}
\usepackage{theorem}
\usepackage{booktabs}

\newtheorem{lemma}{Lemma}
\newtheorem{theorem}{Theorem}
\newtheorem{corollary}{Corollary}

\theoremheaderfont{\sc}\theorembodyfont{\upshape}
\theoremstyle{nonumberplain}
\theoremseparator{}
\theoremsymbol{\rule{1ex}{1ex}}
\newtheorem{proof}{Proof}

\begin{document}

\title{Performance Analysis of $l_0$ Norm Constraint Least Mean Square
Algorithm}

\author{Guolong~Su,~
        Jian~Jin,~
        Yuantao~Gu\thanks{This work was partially supported by National Natural Science Foundation of China (NSFC 60872087 and NSFC U0835003).
The authors are with the Department of Electronic Engineering, Tsinghua University, Beijing 100084, China. The corresponding author of this paper is Yuantao Gu (e-mail: gyt@tsinghua.edu.cn).},~
        and Jian~Wang}

\date{Received June 27, 2011; accepted Dec. 22, 2011.\\\vspace{1em}
This article appears in \textsl{IEEE Transactions on Signal Processing}, 60(5): 2223-2235, 2012.}

\maketitle

\begin{abstract}
As one of the recently proposed algorithms for sparse system
identification, $l_0$ norm constraint Least Mean Square ($l_0$-LMS)
algorithm modifies the cost function of the traditional method with
a penalty of tap-weight sparsity. The performance of $l_0$-LMS is
quite attractive compared with its various precursors. However,
there has been no detailed study of its performance. This paper
presents {comprehensive} theoretical performance analysis
of $l_0$-LMS for white Gaussian input data based on some assumptions
{which are reasonable in a large range of parameter
setting}. Expressions for steady-state mean square deviation
(MSD) are derived and discussed with respect to algorithm parameters
and system sparsity. The parameter selection rule is established for
achieving the best performance. Approximated with Taylor series,
the instantaneous behavior is also derived. In addition, the
relationship between $l_0$-LMS and some previous arts and the
sufficient conditions for $l_0$-LMS to accelerate convergence are set
up. Finally, all of the theoretical results are compared with
simulations and are shown to agree well in a {wide range of parameters}.

\end{abstract}

\section{Introduction}\label{Section:Introduction}

Adaptive filtering has attracted much research
interest in both theoretical and applied issues for a long
time\cite{ref:Book_Widrow,ref:Book_Haykin,ref:Book_Sayed}. Due to
its good performance, easy implementation, and high robustness,
Least Mean Square (LMS) algorithm
\cite{ref:Book_Widrow,ref:Book_Haykin,ref:Book_Sayed,ref:LS_View_SPM}
has been widely used in various applications such as system
identification, channel equalization, and echo cancelation.

The unknown systems to be identified are sparse in most physical
scenarios, including the echo paths \cite{ref:PNLMS1} and Digital TV
transmission channels \cite{ref:Example_DTV}. In other words, there
are only a small number of non-zero entries in the long impulse
response. For such systems, the traditional LMS has no particular
gain since it never takes advantage of the prior sparsity knowledge.
In recent years, several new algorithms have been proposed based on
LMS to utilize the feature of sparsity. M-Max Normalized LMS
(MMax-NLMS)\cite{ref:SLMS} and Sequential Partial Update LMS
(S-LMS)\cite{ref:PULMS} decrease the computational cost and
steady-state mean squared error (MSE) by means of updating filter
tap-weights selectively. Proportionate NLMS (PNLMS) and its improved
version\cite{ref:PNLMS1,ref:PNLMS2} accelerate the convergence by
setting the individual step size in proportional to the respective
filter weights.

Sparsity in adaptive filtering framework has been a long discussed topic \cite{ref:Adaptive_Sparse_1,ref:Adaptive_Sparse_2}.
Inspired by the recently appeared sparse signal processing branch
\cite{ref:l0LMS,ref:SparseLMS,
ref:RLS_l1_1,ref:RLS_l1_2,ref:Adaptive_Sparse_3,ref:Adaptive_Sparse_4,ref:REV_1_Paper_1,ref:REV_2_Paper_4,ref:REV_3_Paper_1},
especially compressive sampling (or compressive sensing,
CS)\cite{ref:CS_Donoho,ref:CS_Tao,ref:CS_Candes}, a family of sparse
system identification algorithms has been proposed based on $l_p$
norm constraint. The basic idea of such algorithms is to exploit the
characteristics of unknown impulse response and to exert sparsity
constraint on the cost function of gradient descent. Specially,
ZA-LMS \cite{ref:SparseLMS} utilizes $l_1$ norm and draws the
zero-point attraction to all tap-weights.
$l_0$-LMS \cite{ref:l0LMS} employs a non-convex approximation of $l_0$ norm and exerts respective
attractions to zero and non-zero coefficients. The smoothed $l_0$ algorithm, which is also based on an approximation of
$l_0$ norm, is proposed in \cite{ref:REV_2_Paper_1} and analyzed in \cite{ref:REV_2_Paper_2}.
Besides LMS variants, RLS-based sparse algorithms \cite{ref:RLS_l1_1,ref:RLS_l1_2} and
Bayesian-based sparse algorithms \cite{ref:REV_2_Paper_3} have also been proposed.

It is necessary to conduct a theoretical analysis for $l_0$-LMS
algorithm. Numerical simulations demonstrate that the mentioned
algorithm has rather good performance compared with several
available sparse system identification algorithms \cite{ref:l0LMS},
including both accelerating the convergence and decreasing the
steady-state MSD. $l_0$-LMS performs \textit{zero-point attraction}
to small adaptive taps and pulls them toward the origin, which
consequently increases their convergence speed and decreases their
steady-state bias. Because most coefficients of a sparse system are
zero, the overall identification performance is enhanced. It is also
found that the performance of $l_0$-LMS is highly affected by the
predefined parameters. Improper parameter setting could not only
make the algorithm less efficient, but also yield steady-state
misalignment even larger than the traditional algorithm. The
importance of such analysis should be further emphasized
since adaptive filter framework and $l_0$-LMS behave well in the solution of sparse signal recovery
problem in compressive sensing\cite{ref:ADF_CS}. Compared with some
convex relaxation methods and greedy pursuits
\cite{ref:CS_IRLS,ref:CS_SpaRSA,ref:CS_OMP}, it was experimentally
demonstrated that $l_0$-LMS in adaptive filtering framework shows
more robustness against noise, requires fewer measurements for
perfect reconstruction, and recovers signal with less sparsity.
Considering its importance as mentioned above, the steady-state
performance and instantaneous behavior of $l_0$-LMS are throughout
analyzed in this work.

\subsection{Main contribution}

One contribution of this work is on steady-state performance
analysis. Because of the nonlinearity
caused by the sparsity constraint in $l_0$-LMS, the theoretical analysis is rather
difficult. To tackle this problem and enable mathematical
tractability, adaptive tap-weights are sorted into different
categories and several assumptions besides the popular independence
assumption are employed. Then, the stability condition on step size
and steady-state misalignment are derived.
After that, the parameter selection
rule for optimal steady-state performance is proposed. Finally, The
steady-state MSD gain is obtained theoretically of $l_0$-LMS over
the tradition algorithm, with the optimal parameter.

Another contribution of this work is on instantaneous behavior analysis, which indicates the
convergence rate of LMS type algorithms and also arouses much attention
\cite{ref:Analysis_LeakyGaussian,ref:Analysis_Deficient,ref:Analysis_MSE_TransientSpeed}. For LMS and most of its linear variants, the
convergence process can be obtained in the same derivation procedure as steady-state misalignment. However, this no longer holds for $l_0$-LMS due to its nonlinearity. In a different way by utilizing the obtained steady-state MSD as foundation, a Taylor expansion is employed to get an
approximated quantitative analysis of the convergence process. Also, the convergence rates are compared between $l_0$-LMS and standard LMS.

\subsection{Relation to other works}
In order to theoretically characterize the performance and guide the
selection of the optimal algorithm parameters, the mean square
analysis has been conducted for standard LMS and a lot of its
variants. To the best of our
knowledge, Widrow for the first time proposed the LMS algorithm in
\cite{ref:Analysis_Widrow_1960} and studied its performance in
\cite{ref:Analysis_Widrow_1976}. Later, Horowitz and Senne
\cite{ref:Analysis_ComplexNarrowBand} established the mathematical
framework for mean square analysis via studying the weight vector
covariance matrix and achieved the closed-form expression of MSE,
which was further simplified by Feuer and Weinstein
\cite{ref:Analysis_ConvergenceWhiteGaussian}. The mean square
performance of two variants, leaky LMS and deficient length LMS,
were theoretically investigated in similar methodologies in
\cite{ref:Analysis_LeakyGaussian} and \cite{ref:Analysis_Deficient},
respectively. Recently, Dabeer and Masry
\cite{ref:Analysis_MSE_TransientSpeed} put forward a new approach
for performance analysis on LMS without assuming a linear regression
model. Moreover, convergence behavior of transform-domain LMS was
studied in \cite{ref:Analysis_TransformLMS} with second-order
autoregressive process.  A summarized analysis was proposed in
\cite{ref:Analysis_TimeInvariantGains} on a class of adaptive
algorithms, which performs linear time-invariant operations on the
instantaneous gradient vector and includes LMS as the simplest case.
Similarly, the analysis of Normalized LMS has also attracted much
attention \cite{ref:Analysis_Conv_NLMS_1,ref:Analysis_Conv_NLMS_2}.

However, the methodologies mentioned above, which are effective in
their respective context, could no longer be directly applied to
the analysis of $l_0$-LMS, considering its high non-linearity.
Admittedly, nonlinearity is a long topic in adaptive filtering and
not unique for $l_0$-LMS itself. Researchers have delved into the
analysis of many other LMS-based nonlinear variants
\cite{ref:Analysis_l2lp,ref:Analysis_ErrorNonlinearity,ref:Analysis_Steady_ErrorNonlinearity,ref:Analysis_Least_MeanM,ref:ShiKun,ref:Analysis_SignSign_1,ref:Analysis_SignSign_2,ref:Analysis_SignSign_3,ref:Analysis_SignSign_4}.
Nevertheless, the nonlinearity of most above references comes from
non-linear operations on \emph{the estimated error}, rather than
\emph{the adaptive tap-weights} that $l_0$-LMS mainly
focuses on.

We have noticed that the mean square deviation analysis of ZA-LMS has been conducted
\cite{ref:ShiKun}. However, this work is far different from the reference. First of all, the
literature did not consider the transient performance analysis while in this work the mean square
behavior of both steady-state and convergence process are conducted. Moreover, considering
$l_0$-LMS is more sophisticated than ZA-LMS, there are more parameters in $l_0$-LMS than in ZA-LMS,
which enhances the algorithm performance but increases the difficulty of theoretical analysis. Last
but not least, taking its parameters to a specific limit setting, $l_0$-LMS becomes essentially the same as ZA-LMS, which can apply the theoretical results of this work directly.

A preliminary version of this work has been presented in
conference\cite{ref:Analysis_Self}, including the convergence
condition, derivation of steady-state MSD, and an expression of the
optimal parameter selection. This work provides not only a
detailed derivation for steady-state results, but also the mean
square convergence analysis. Moreover,
both the steady-state MSD and the
parameter selection rule are further simplified and available for
analysis. Finally, more simulations are performed to validate the
results and more discussions are conducted.

This paper is organized as follows. In section \ref{Section:Review},
a brief review of $l_0$-LMS and ZA-LMS is presented. Then
in section \ref{Section:Preliminaries}, {a few assumptions
are introduced which are reasonable in a wide range of situations}. Based on these assumptions, section
\ref{Section:Analysis} proposes the mean square
analysis. Numerical experiments are performed to demonstrate the
theoretical derivation in section \ref{Section:Simulation} and the
conclusion is drawn in section \ref{Section:Conclusion}.

\section{Background}\label{Section:Review}

\subsection{$l_0$-LMS algorithm}

The unknown coefficients and input signal at time
instant $n$ are denoted by ${\bf s}=\left[s_0, s_1,\ldots,
s_{L-1}\right]^{\rm T}$ and ${\bf x}_n=\left[x_n,x_{n-1},\cdots,
x_{n-L+1}\right]^{\rm T}$, respectively, where $L$ is the filter
length. The observed output signal is
\begin{equation}
d_{n} = {\bf x}^{\rm T}_{n} {\bf s} + v_n, \label{ExpOut}
\end{equation}
where $v_n$ denotes the additive noise. The estimated error between the output of unknown system and of the adaptive filter is
\begin{equation}
e_n=d_n-{\bf x}^{\rm T}_n {\bf w}_n, \label{EstError}
\end{equation}
where ${\bf w}_n=\left[w_{0,n},w_{1,n},\cdots, w_{L-1,n}\right]^{\rm T}$ denotes the adaptive filter tap-weights.

In order to take the sparsity of the unknown coefficients into
account, $l_0$-LMS \cite{ref:l0LMS} inserts an $l_0$ norm penalty
into the cost function of standard LMS. The new cost function is
$$
\xi_n= e^2_n+\gamma \|{\bf w}_n\|_0,
$$
where $\gamma>0$ is a factor to balance the estimation error and the
new penalty. Due to the NP hardness of $l_0$ norm optimization, a
 continuous function is usually
employed to approximate $l_0$ norm. Taking the popular
approximation\cite{ref:l0_Approximation} and
making use of the first order Taylor expansion,
the recursion of $l_0$-LMS is
\begin{equation}
{\bf w}_{n+1}={\bf w}_n+\mu e_n {\bf x}_n + \kappa g({\bf w}_n),\label{w_Update}
\end{equation}
where $g({\bf w}_n)=\left[g(w_{0,n}),g(w_{1,n}),\cdots,g(w_{L-1,n})\right]^{\rm
T}$ and
\begin{equation}
\label{l_0GradApprx} g(t)= \begin{cases}
2\alpha^2 t-2\alpha\cdot{\rm sgn}(t) & |t|\le 1/\alpha;\\
0 &  {\rm elsewhere}. \end{cases}
\end{equation}
The last item in (\ref{w_Update}) is called \emph{zero-point
attraction} \cite{ref:l0LMS}, \cite{ref:ADF_CS}, because it reduces
the distance between $w_{i,n}$ and the origin when $|w_{i,n}|$ is
small. According to (\ref{l_0GradApprx}) and Fig.
\ref{Fig:g_l0l1}(a), obviously such {attraction} is non-linear and
exerts varied affects on respective tap-weights. This {attraction} is
effective for the tap-weights in the interval $\left(-{1}/{\alpha},
{1}/{\alpha}\right)$, which is named \emph{attraction range}. In
this region, the smaller $|w_{i,n}|$ is, the stronger attraction
affects.

\subsection{ZA-LMS and RZA-LMS}
\begin{figure}
  \centering
  \includegraphics[width=3.2in]{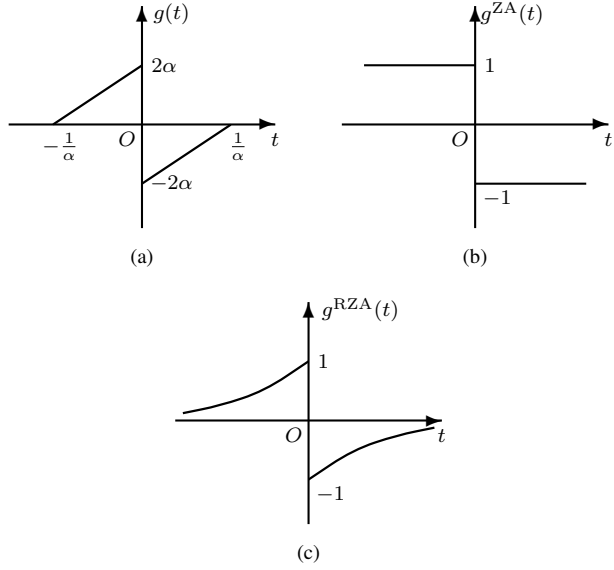}
  \caption{The {Zero}-point attraction of (a) $l_0$-LMS, (b) ZA-LMS, (c) RZA-LMS.}\label{Fig:g_l0l1}
\end{figure}

ZA-LMS (or Sparse LMS) \cite{ref:SparseLMS} runs similar as
$l_0$-LMS. The only difference is that the sparse penalty is changed
to $l_1$ norm. Accordingly the zero-point attraction item of the
former is defined as
\begin{equation}
\label{l_1GradApprx} g^{\rm ZA}(t)= -{\rm sgn}(t),
\end{equation}
which is shown in Fig. \ref{Fig:g_l0l1}(b).
The recursion of ZA-LMS is
\begin{equation}
\label{w_Update_l1}{\bf w}_{n+1}={\bf w}_n+\mu e_n {\bf x}_n + \rho g^{\rm ZA}({\bf x}_n),
\end{equation}
where $\rho$ is the parameter to control the strength of sparsity penalty. Comparing
the sub figures in Fig. \ref{Fig:g_l0l1}, one can readily accept that $g(t)$ exerts the various
attraction to respective tap-weight, therefore it usually behaves better than $g^{\rm
ZA}(t)$. In the following analysis, one will read that ZA-LMS is a special case of $l_0$-LMS and
the result of this work can be easily extended to the case of ZA-LMS.

As its improvement, Reweighted ZA-LMS (RZA-LMS) is also proposed in \cite{ref:SparseLMS},
which modifies the zero-point attraction term to
\begin{equation}
\label{RZAGradApprx} g^{\rm RZA}(t)= -\frac{{\rm sgn}(t)}{1+\varepsilon |t|},
\end{equation}
where parameter $\varepsilon$ controls the similarity between (\ref{RZAGradApprx}) and $l_0$ norm.
Please refer to Fig. \ref{Fig:g_l0l1}(c) for better understanding the behavior of (\ref{RZAGradApprx}).
In section V, both ZA-LMS and RZA-LMS are simulated for the purpose of performance comparison.

\subsection{Previous results on LMS and ZA-LMS}

Denote $D^{\rm LMS}_\infty$ and $D^{\rm LMS}_n$ as the steady-state MSD
and instantaneous MSD after $n$ iterations for LMS  with
zero-mean independent Gaussian input, respectively. The steady-state MSD has the
explicit expression \cite{ref:Book_Haykin} of
\begin{equation}\label{MSDLMS}
    D^{\rm LMS}_\infty=\frac{\mu P_vL}{2-\mu P_x(L+2)}
    =\frac{\mu P_vL}{\Delta_L},
\end{equation}
where $P_x$ and $P_v$ denote the power of input signal and additive noise, respectively, and
$\Delta_L$ is a constant defined by (\ref{DefineDeltaL}) in Appendix \ref{Const}.
For the convergence process, the explicit expression of instantaneous MSD is implied in \cite{ref:Analysis_ComplexNarrowBand} as
\begin{equation}\label{D_Conv_ClosedForm_LMS}
    D^{\rm LMS}_n=\frac{\mu P_vL}{\Delta_L}
    +\left(\|{\bf s}\|_2^2 -\frac{\mu P_vL}{\Delta_L}\right)
    \cdot\left(1-\mu P_x\Delta_L\right)^n.
\end{equation}

The next one turns to ZA-LMS, $D^{\rm ZA}_\infty$ is used to denote the steady-state MSD with white Gaussian input. Reference
\cite{ref:ShiKun} reaches the conclusion that
\begin{equation}\label{Shi_Sparse_LMS_MSD}
    D^{\rm ZA}_\infty=\frac{2}{\mu P_x}\left(y^2-\frac{\pi\mu P_x+\Delta_0}{2\pi \mu^2 P_x^2}
    \rho^2\right)-\frac{P_v}{P_x},
\end{equation}
where $y$ is the solution to
\begin{align*}
    \Delta_L y^2+(L-Q)\rho\sqrt{\frac{2\Delta_0}{\pi}}y
    -\left(\frac{L-2Q}{2\pi}+Q+1\right)\frac{\Delta_0\rho^2}{\mu P_x} -\frac{\Delta_0^2\rho^2}{\pi\mu^2 P_x^2}
    -\mu P_v\Delta_0=0,\label{Shi_Sparse_LMS_rho}
\end{align*}
where $Q\le L$ denotes the number of non-zero unknown coefficients
and $\Delta_0$ is a constant defined by (\ref{DefineDelta0}).

\subsection{Related steepest ascent algorithms for sparse decomposition}

$l_0$-LMS employs steepest descent recursively and is applicable to solving sparse system identification. More generally, steepest ascent iterations are used in several algorithms in the field of sparse signal processing. For example, researchers developed smoothed $l_0$ method \cite{ref:REV_2_Paper_1} for sparse decomposition, whose iteration includes a steepest ascent step and a projection step. The first step is defined as
\begin{equation}
\label{SL0_1}
{\bf \hat{w}}_{n+1}={\bf w}_{n}+\mu {\mathbf{v}}^{\rm SL0}_{n},
\end{equation}
where $\mu$ serves as step size, ${\mathbf{v}}^{\rm SL0}_{n}=[v^{\rm SL0}_{0,n},v^{\rm SL0}_{1,n},...,v^{\rm SL0}_{L-1,n}]^{\rm T}$ denotes the negative derivative to an approximated $l_0$ norm and takes the value
\begin{equation}
\label{SL0_2}
v^{\rm SL0}_{k,n}=-w_{k,n}{\rm exp}\left(-2w_{k,n}^2/\sigma^2\right),\ \ 0\le k<L.\nonumber
\end{equation}
After (\ref{SL0_1}), a projection step is performed which maps ${\bf \hat{w}}_{n+1}$ to ${\bf w}_{n+1}$ in the feasible set. It can be seen that (\ref{SL0_1}) performs steepest ascent, which is similar to zero-point attraction in $l_0$-LMS. The iteration details and performance analysis of this algorithm are presented in \cite{ref:REV_2_Paper_1} and \cite{ref:REV_2_Paper_2}, respectively.

Another algorithm, named Iterative Bayesian\cite{ref:REV_2_Paper_3}, also enjoys steepest ascent iteration as
\begin{equation}
\label{IBay}
{\bf w}_{n+1}={\bf w}_{n}+\mu\frac{\partial L}{\partial {\bf w}},
\end{equation}
where $\mu$ denotes the step size and $L$ is a log posterior probability function.
Analysis of this algorithm and its application to sparse component analysis in noisy scenario are presented in \cite{ref:REV_2_Paper_3}.

\section{Preliminaries}\label{Section:Preliminaries}

Considering the nonlinearity of zero-point attraction, some
preparations are made to simplify the mean square performance analysis.

\subsection{Classification of unknown coefficients}

Because various affects are exerted in $l_0$-LMS to the filter tap-weights
according to their respective system coefficients, it would be
helpful to classify the unknown parameters, correspondingly, the
filter tap-weights, into several categories and perform different analysis on
each category separately. According to the attraction range and their strength,
all system coefficients are classified into three
categories as
\begin{eqnarray}
{\rm Large\ coefficients}:&& \mathcal{C}_L=\left\{k{\big|}|s_k|\ge 1/\alpha\right\};\nonumber\\
{\rm Small\ coefficients}:
&&\mathcal{C}_S=\left\{k{\big|}0<|s_k|<1/\alpha\right\};\nonumber\\
{\rm Zero\ coefficients}:
&&\mathcal{C}_0=\left\{k{\big|}s_k=0\right\},\nonumber\label{Classification_LCSCZC}
\end{eqnarray}
where $0\le k<L$. Obviously, $|\mathcal{C}_L\cup \mathcal{C}_S\cup \mathcal{C}_0| = L$ and $|\mathcal{C}_L\cup \mathcal{C}_S| = Q$. In the following text, derivations are firstly
carried out for the three sets separately. Then a synthesis is taken to achieve the final
results.

\subsection{Basic assumptions}

The following assumptions about the
system and the predefined parameters are adopted to enable the formulation.

\begin{itemize}
\item[{(i)}]
Input data $x(n)$ is an \emph{i.i.d.} zero-mean Gaussian signal.

\item[{(ii)}]
Tap-weights ${\bf w}_n$, input vector ${\bf x}_n$, and additive
noise $v_n$ are mutually independent.

\item[{(iii)}]
The parameter $\kappa$ is so small that $2\alpha^2\kappa\ll\mu P_x$.
\end{itemize}

Assumption (i) commonly holds while (ii) is the well-known
independence assumption \cite{ref:Book_Sayed}. Assumption (iii) comes from the
experimental observations, i.e., a too large $\kappa$ can cause much bias as well as large steady-state MSD. Therefore, in order to achieve better
performance, $\kappa$ should not be too large.

Besides the above items, several regular patterns are supposed during the convergence {and the steady state}.

\begin{itemize}
\item[{(iv)}]
All tap-weights, ${\bf w}_n$, follow Gaussian distribution.

\item[{(v)}]For $k\in \mathcal{C}_L\bigcup \mathcal{C}_S$, the tap-weight
$w_{k,n}$ is assumed to have the same sign with the corresponding
unknown coefficient.

\item[{(vi)}]The adaptive weight $w_{k,n}$ is assumed out of the
attraction range for $k\in \mathcal{C}_L$, while in the attraction range elsewhere.
\end{itemize}

Assumption (iv) is usually accepted for steady-state behavior
analysis \cite{ref:SparseLMS,ref:Analysis_SignSign_3}. {Assumption (v) and (vi) are
considered suitable in this work due to the following} two aspects.
First, there are few taps violating these assumptions in a common
scenario. Intuitively, only the non-zero taps with rather small
absolute value may violate assumption (v), while assumption (vi) may
not hold for the taps close to the boundaries of the attraction
range. For other taps which make up the majority, these assumptions are usually
reasonable, especially in high SNR cases. Second, assumptions (v) and (vi) are proper for
small steady-state MSD, which is emphasized in this work. The
smaller steady-state MSD is, the less tap-weights differ from
unknown coefficients. Therefore, it is more likely that they share
the same sign, as well as on the same side of the attraction range.

Based on the discussions above, those patterns {are regarded suitable} in steady state. For the convergence process, due to fast convergence of LMS-type algorithms, {we may suppose that most taps will get
close to the corresponding unknown coefficients very quickly, so these patterns are also employed} in common scenarios. As we will see later, some of the above assumptions cannot always
hold in whatever parameter setting and may restrict the applicability of some analysis below.
However, considering the difficulties of nonlinear algorithm performance analysis, these
assumptions can significantly enable mathematical tractability and help obtain results shown to be
precious in a large range of parameter setting. Thus, we consider these assumptions reasonable to
be employed in this work.

\section{Performance analysis}\label{Section:Analysis}

Based on the assumptions above, the mean and mean-square performances of $l_0$-LMS are analyzed in this section.

\subsection{Mean performance}
Define the misalignment vector as ${\bf h}_n={\bf w}_n-{\bf s}$,
combine (\ref{ExpOut}), (\ref{EstError}), and (\ref{w_Update}), one
has
\begin{equation}
{\bf h}_{n+1}=\left({\bf I}-\mu {\bf x}_n {\bf x}^{\rm T}_n
\right){\bf h}_n+\mu v_n {\bf x}_n+\kappa g({\bf w}_n).
\label{h_Update}
\end{equation}
Taking expectation and using the assumption (ii), one derives
$$
\overline{{\bf h}_\infty} = \frac{\kappa}{\mu P_x} \overline{g({\bf
w}_\infty)},\label{h_SteadyState_General}
$$
where \emph{overline} denotes expectation.
\begin{itemize}
\item
For $k\in \mathcal{C}_L$, utilizing assumption (vi), one has $\overline{g(w_{k,\infty})}=0$.
\item
For $k\in \mathcal{C}_S$, combining assumptions (iii), (v) and (vi), it can be derived that
$$
\overline{h_{k,\infty}}=\left(1-\frac{2\alpha^2\kappa}{\mu P_x}\right)\frac{\kappa g(s_k)}{\mu P_x}\\
\approx\frac{\kappa g(s_k)}{\mu P_x}.
$$
\item
For $k\in \mathcal{C}_0$, noticing the fact that $g(x)$ has the opposite sign with $x$ in interval $(-1/\alpha, 1/\alpha)$ and using assumptions (iv) and (vi), it can be derived that $\overline{g(w_{k,\infty})}=0$.
\end{itemize}
Thus, the bias in steady
state is obtained
\begin{equation}
\label{h_SteadyState_Detail} \overline{h_{k,\infty}}=
\begin{cases}
\frac{\kappa g(s_k)}{\mu P_x} & k\in \mathcal{C}_S;\\
0 & {\rm elsewhere}.
\end{cases}
\end{equation}
In steady state, therefore, the tap-weights are unbiased for large
coefficients and zero coefficients, while they are biased for small
coefficients. The misalignment depends on the predefined
parameters as well as unknown coefficient $s_k$ itself. The smaller
the unknown coefficient is, the larger the bias becomes. This tendency can be
directly read from Fig. \ref{Fig:g_l0l1}(a). In the attraction
range, the intensity of the zero-point attraction increases as
tap-weights get more closing to
zero, which causes heavy bias. Thus, the bias of small coefficients
in steady state is the byproduct of the attraction, which
accelerates the convergence rate and increases steady-state MSD.

\subsection{Mean square steady-state performance}

The condition on mean square convergence and steady-state MSD are given by the following theorem.

\begin{theorem}
\label{Theo1} In order
to guarantee convergence, step-size $\mu$ should satisfy
\begin{equation}
0<\mu<\mu_{\rm max} = \frac{2}{(L+2)P_x},\label{ConvCondition}
\end{equation}
and the final mean square deviation of $l_0$-LMS is
\begin{equation}
D_\infty = \frac{\mu P_v L}{\Delta_L}+\beta_1 \kappa^2
-\beta_2 \kappa \sqrt{\kappa^2+\beta_3}, \label{MSD_Expression}
\end{equation}
where $\{\beta_i\}$ are defined in
(\ref{beta1})$\sim$(\ref{beta3}) in Appendix \ref{Const},
respectively.
\end{theorem}

The proof of Theorem \ref{Theo1} goes in Appendix
\ref{Proof_Steady_MSD}.

{\it Remark 1}: The steady-state MSD of $l_0$-LMS is composed of two parts: the first item in (\ref{MSD_Expression}) is
exactly the steady-state MSD of standard LMS (\ref{MSDLMS}), while the latter two items compose an additional part caused by zero-point
attraction. When $\kappa$ equals zero, $l_0$-LMS becomes the traditional LMS, and correspondingly the additional part vanishes. When the additional part is negative, $l_0$-LMS has smaller steady-state MSD and thus better steady-state performance over standard LMS. Consequently, it can be deduced that the
condition on $\kappa$ to ensure $l_0$-LMS outperforms LMS in steady-state is
$$
0<\kappa<\frac{\beta_2^2\beta_3}{\beta_1^2-\beta_2^2}.
$$

{\it Remark 2}: According to Theorem \ref{Theo1}, the following corollary on parameter $\kappa$
is derived.
\begin{corollary}\label{Coro:Opt_Kappa}
From the perspective of steady-state performance, the best choice
for $\kappa$ is
\begin{equation}
\kappa_{\rm {opt}}= \frac{\sqrt{
\beta_3}}{2}\left(\sqrt[4]{\frac{\beta_1+\beta_2}{\beta_1-\beta_2}}-\sqrt[4]{\frac{\beta_1-\beta_2}{\beta_1+\beta_2}}\right),
\label{KappaBest}
\end{equation}
and the minimum steady-state MSD is
\begin{equation}
D^{\rm min}_\infty = \frac{\mu P_v L}{\Delta_L}+\frac{
\beta_3}{2}\left(\sqrt{\beta_1^2-\beta_2^2}-\beta_1\right).
\label{MSD_Opt}
\end{equation}
\end{corollary}

The proof of Corollary $\ref{Coro:Opt_Kappa}$ is presented in
Appendix \ref{Best_Parameters}. Please notice that in
(\ref{MSD_Opt}), the first item is about standard LMS and the second
one is negative when $Q$ is less than $L$.
Therefore, the minimum steady-state MSD of $l_0$-LMS is less than
that of standard LMS as long as the
system is not totally non-sparse.

{\it Remark 3}:
 According to the
theorem, it can be accepted that the steady-state MSD is not only
controlled by the predefined parameters, but also dependent on
the unknown system in the following two aspects. First, the sparsity of the system response, i.e. $Q$ and $L$,
controls the steady-state MSD. Second, significantly
different from standard LMS, the steady-state MSD is relevant to the
\emph{small coefficients} of the system, considering the attracting strength appears in $\beta_0$ and $\beta_1$.

Here we mainly discuss the effect of system sparsity as well as the distribution of coefficients on the minimum steady-state MSD. Based on
the above results, the following corollary can be deduced.

\begin{corollary}\label{Coro:Steady_Q}
The minimum steady-state MSD of (\ref{MSD_Opt}) is monotonic
increasing with respect to $Q$ and attracting strength $G({\bf s})$.
\end{corollary}

The validation of Corollary $\ref{Coro:Steady_Q}$ is performed in
Appendix \ref{Effect_Sparsity}. The \emph{zero-point
attractor} is utilized in $l_0$-LMS to draw tap-weights towards zero.
Consequently, the more sparse the unknown system is, the less
steady-state MSD is. Similarly, small
coefficients are biased in steady state and deteriorate the
performance, which explains that steady-state
MSD is increasing with respect to $G({\bf s})$.

{\it Remark 4}:
According to (\ref{ConvCondition}), one knows that $l_0$-LMS has the same convergence condition on step size as standard LMS\cite{ref:Book_Haykin} and ZA-LMS\cite{ref:ShiKun}. Consequently the effect of $\mu$ on steady-state performance is analyzed.
It is indicated in
(\ref{MSDLMS}) that the standard LMS enhances steady-state
performance by reducing step size\cite{ref:Book_Haykin}. $l_0$-LMS
has a similar trend.
For the seek of simplicity and practicability, a sparse system of $Q$ far less than $L$ is considered to demonstrate this property.
Utilizing (\ref{ConvCondition}) in such scenario, the following corollary is derived.

\begin{corollary}\label{Coro:Steady_mu}
For a sparse system which satisfies
\begin{equation}
Q\ll L\quad {\rm and}\quad (Q+2)\mu P_x\ll
2,\label{Sparse_Condition}
\end{equation}
the minimum
steady-state MSD in (\ref{MSD_Opt}) is further approximately
simplified as
\begin{equation}
D^{\rm min}_\infty\approx
\frac{\mu P_v L}{\Delta_L}\!\!
\left(\!1\!-\!\frac{\eta_6}{\eta_5+\eta_6+\!\!\sqrt{\eta_5^2+\frac{32\alpha^2
L}{\pi}G({\bf s})}}\right)\!\!,\label{MSD_Simple_Sparse_SmallMu}
\end{equation}
where $\eta_5$ and $\eta_6$ are defined by (\ref{eta6}) in Appendix
\ref{Const}, and $G({\bf s})$, defined by
(\ref{AttractingStrength}), denotes the attracting strength
to the zero-point. Furthermore, the minimum steady-state MSD increases with
respect to the step size.
\end{corollary}

The proof of Corollary \ref{Coro:Steady_mu} is conducted in Appendix \ref{Effect_Mu}. Due to the stochastic gradient descent and zero-point attraction, the
tap-weights suffer oscillation, even in steady state, whose intensity is directly relevant to the step size. The larger the step size, the more intense the vibration. Thus, the steady-state MSD is monotonic increasing with respect to $\mu$ in the above scenario.

{\it Remark 5}: In the scenario where $2\alpha\kappa=\rho$ remains a constant while $\alpha$
approaches to zero, it can be readily accepted that (\ref{w_Update}) becomes totally identical to
(\ref{w_Update_l1}), therefore $l_0$-LMS becomes ZA-LMS in this limit setting of parameters. In
Appendix \ref{Relationship_Sparse_Standard_LMS} it is shown that the result
(\ref{Shi_Sparse_LMS_MSD}) for steady-state performance \cite{ref:SparseLMS} could be regarded as a
particular case of Theorem \ref{Theo1}. As $\alpha$ approaches to zero in $l_0$-LMS, the attraction
range becomes infinity and all non-zero taps belong to small coefficients which are biased in
steady state. Thus, ZA-LMS has larger steady-state MSD than $l_0$-LMS, due to bias of all taps
caused by uniform attraction intensity. If $\kappa$ is further chosen optimal, the optimal
parameter for ZA-LMS is given by $\rho_{\rm {opt}}=\lim_{\alpha\rightarrow 0}2\alpha\kappa_{\rm
{opt}}$ (notice that $\kappa_{\rm {opt}}$ approaches $\infty$ as $\alpha$ tending to zero, as
makes $\rho_{\rm {opt}}$ finite), and the minimum steady-state MSD of $l_0$-LMS (\ref{MSD_Opt})
converges to that of ZA-LMS. To better compare the three algorithms, the steady-state MSDs of LMS, ZA-LMS, and $l_0$-LMS are listed in TABLE \ref{table1}, where that of ZA-LMS is rewritten and $\Gamma$ is defined in (\ref{Gamma_ZA}) in Appendix \ref{Relationship_Sparse_Standard_LMS}.
It can be accepted that the steady-state MSDs of both ZA-LMS and $l_0$-LMS are in the form of $D_{\infty}^{\rm LMS}$ plus addition items, where $D_{\infty}^{\rm LMS}$ denotes the steady-state MSD of standard LMS. If the additional items are negative, ZA-LMS and $l_0$-LMS exceed LMS in steady-state performance.
\begin{table*}[!t]
\begin{center}
\begin{scriptsize}
\caption{The steady-state MSDs of three algorithms.} \label{table1}
\begin{scriptsize}
\begin{tabular}{ccccc}
\hline
 &  & \multicolumn{3}{c}{Steady-state MSD} \\ \cline{3-5}
\raisebox{1.3ex}[0pt]{Alg.} & \raisebox{1.3ex}[0pt]{Relat. w. $l_0$-LMS} & Eq. No. & Denota. & Expression \\ \hline \vspace{0.5em}
$l_0$-LMS & --- & (\ref{MSD_Expression}) & $D_\infty$ & $D^{\rm LMS}_\infty+\beta_1 \kappa^2
-\beta_2 \kappa \sqrt{\kappa^2+\beta_3}$\\ \vspace{0.5em}
ZA-LMS &  $2\alpha\kappa\!=\!\rho\!$ and $\!\alpha\!\rightarrow\!0$  & (\ref{MSD_L1_Expression}) & $D^{\rm ZA}_\infty$ & $D^{\rm LMS}_\infty\!-\!\frac{\rho(L-Q)\sqrt{\Gamma}}{\sqrt{2\pi}\mu^2 P_x^2 \Delta_L^2}\!+\!\frac{\rho^2(2(L-Q)\Delta_0\Delta_Q+\pi \Delta_L (\mu LP_x + 2Q\Delta_0))}{\pi\mu^2 P_x^2 \Delta_L^2}$\\ \vspace{0.5em}
LMS & $\kappa=0$ & (\ref{MSDLMS}) & $D^{\rm LMS}_\infty$ & $\frac{\mu P_vL}{\Delta_L}$\\ \hline
\end{tabular}
\end{scriptsize}
\end{scriptsize}
\end{center}
\end{table*}

{\it Remark 6}: Now the extreme case that all taps in system are zero, i.e. $Q=0$, is
considered. If $\kappa$ is set as the optimal, (\ref{MSD_Opt}) becomes
\begin{equation}
\label{MSD_Q0}
D_{\infty}^{\rm min}=\frac{\mu P_v L}{\Delta_L}-\frac{2\mu P_v L \Delta_0^2}{2\Delta_L\Delta_0^2+\pi\mu P_x \Delta_L^2}.
\end{equation}
Due to the independence of (\ref{MSD_Q0}) on $\alpha$, this result also holds in the scenario of
$\alpha$ approaching zero; thus, (\ref{MSD_Q0}) also applies for the steady-state MSD of ZA-LMS with optimal $\rho$, in the extreme case $Q=0$. Thus, it has
been shown that $l_0$-LMS and ZA-LMS with respective optimal parameters have the same steady state
performance for a system with all coefficients zero. Although this result seems a little strange at
the first sight, it is in accordance with intuition considering the zero-point attraction item in
$l_0$-LMS. Since the system only has zero taps, all $w_{k,\infty}$ only vibrate in a very small
region around zero. The zero-point attraction item is $\kappa g(t) \approx -2\alpha\kappa {\rm
sgn}(t)$ when $t$ is very near zero, thus as long as we set $\alpha\kappa$ to be constant, the item mentioned above
and the steady state MSD have little dependence on $\alpha$ itself. Thus, when
$\kappa$ is chosen as optimal and $Q=0$, the steady state MSD generally does not change with
respect to $\alpha$.

\subsection{Mean square convergence behavior}
Based on the results achieved in steady state, the convergence process can
be derived approximately.
\begin{lemma} \label{lemma}
The instantaneous MSD is the solution to the first order
difference equations
\begin{equation}
\left[\!\!
\begin{array}{c}
D_{n+1}\\
\Omega_{n+1}
\end{array}\!\!
\right] ={\mathbf{A}} \left[\!\!
\begin{array}{c}
D_n\\
\Omega_n
\end{array}\!\!
\right]+{\mathbf{b}_n}
 ,
  \label{D_Omega_Iter}
\end{equation}
where $\Omega_n = \sum_{k\in
\mathcal{C}_0}\overline{h^2_{k,n}}$, vector ${\bf b}_n$ and constant
matrix ${\bf A}$ are defined in (\ref{Vector_b}) and
(\ref{Matrix_A}) in Appendix \ref{Const}, respectively. Initial
values are
\begin{equation}
\left[\!\!
\begin{array}{c}
D_0\\
\Omega_0
\end{array}\!\!
\right] = \left[\!\!
\begin{array}{c}
\|{\bf s}\|_2^2\\
0
\end{array}\!\!
\right].
  \label{D_Omega_Init}
\end{equation}
\end{lemma}

The derivation of Lemma \ref{lemma} goes in Appendix
\ref{Proof_Instantaneous_MSD}. Since $\omega$, which is defined by
(\ref{Quadratic_EQU}), appears in both $\bf A$ and ${\bf b}_n$, the
convergence process is affected by algorithm parameters, the length
of system, the number of
non-zero unknown coefficients, and the strength or distribution of
\emph{small coefficients}. Moreover, derivation in Appendix
\ref{lemma_To_Theo2} yields the solution to (\ref{D_Omega_Iter}) in
the following theorem.
\begin{theorem}
\label{Theo2} The closed form of instantaneous MSD is
\begin{equation}\label{D_Conv_ClosedForm}
D_n = c_1\lambda_1^n+c_2\lambda_2^n+c_3\lambda_3^n+D_\infty,
\end{equation}
where $\lambda_1$ and $\lambda_2$ are the eigenvalues of matrix
${\mathbf{A}}$, $c_1$ and $c_2$ are coefficients defined by initial
values (\ref{D_Omega_Init}). The expressions of constants
$\lambda_3$ and $c_3$
 are listed in (\ref{lambda3}) and (\ref{C3}) in Appendix
\ref{Const}, respectively. $D_\infty$ denotes the steady-state
MSD.
\end{theorem}

The two eigenvalues can be easily calculated. Through the method of undetermined coefficients, $c_1$
and $c_2$ are obtained by satisfying initial values $D_0$ and $D_1$, which is acquired by (\ref{D_Omega_Iter}) and (\ref{D_Omega_Init}).
Considering the high complexity of their closed form expressions, they are not included in this paper for the sake of simplicity.

Next we discuss the relationship of mean square convergence between
$l_0$-LMS and standard LMS. In the scenario where $l_0$-LMS with zero $\kappa$
 becomes traditional LMS, it can be shown after some calculation that {$c_2=c_3=0$} in (\ref{D_Conv_ClosedForm}), which becomes in
accordance with (\ref{D_Conv_ClosedForm_LMS}). Now we turn to the
MSD convergence rate of these two algorithms. From the perspective
of step size, one has the following corollary.

\begin{corollary}
\label{Coro:Fast_Conv_mu} A
sufficient condition for that $l_0$-LMS finally converges more quickly
than LMS is $\mu_{\rm max}/2<\mu<\mu_{\rm max}$, where $\mu_{\rm max}$ is defined in (\ref{ConvCondition}).
\end{corollary}

The proof is postponed to Appendix \ref{Proof_Coro:Fast_Conv_mu}.
From Corollary \ref{Coro:Fast_Conv_mu}, one knows that for a large
step size, the convergence rate of $l_0$-LMS is finally faster
than that of LMS. However, this condition is not necessary. In fact,
$l_0$-LMS can also have faster convergence rate for small step size,
as shown in numerical simulations.

On the perspective of the system coefficients distribution, one has
another corollary.
\begin{corollary}
\label{Coro:Fast_Conv_SC} Another
sufficient condition to ensure that $l_0$-LMS finally enjoys acceleration is
$$
\mathcal{C}_S=\emptyset,\quad
\text{or equivalently,}\quad
\alpha\ge \max_{s_k\ne 0}\frac{1}{|s_k|}.
$$
\end{corollary}

This corollary is obtained from the
fact that $c_3$ equals zero in this condition, using the similar
proof in Appendix \ref{Proof_Coro:Fast_Conv_mu}. The full
demonstration is omitted to save space. Therefore, for sparse
systems whose most coefficients are exactly zeros, a large enough
$\alpha$ guarantees faster convergence rate finally. Similar as above, this
condition is also not necessary. $l_0$-LMS {can converge} rather fast
even if such condition is violated.

\section{Numerical experiments}\label{Section:Simulation}

Five experiments are designed to confirm the theoretical analysis.
The non-zero coefficients of the unknown system are Gaussian
variables with zero mean and unit variance and their locations are
randomly selected. Input signal and additive
noise are white zero mean Gaussian series with
various signal-to-noise ratio. Simulation
results are the averaged deviation of $100$ independent trials. For
theoretical calculation, the expectation of attracting strength in
(\ref{AttractingStrength}) and (\ref{AttractingStrength1}) are
employed to avoid the dependence on priori knowledge of system. The
parameters of these experiments are listed in TABLE \ref{table2},
where $\kappa_{\rm opt}$ is calculated by (\ref{KappaBest}).

\begin{table*}[!t]
\begin{center}
\begin{scriptsize}
\begin{scriptsize}
\caption{The parameters in experiments.} \label{table2}
\begin{tabular}{ccccccc}
\hline
Experiment & $L$ & $Q$ & $\mu$ & $\alpha$ & $\kappa$ & SNR \\ \hline
1 & 1000 & 100  & $8$E$-4$ & $10$ & $1$E$-9\rightarrow 3$E$-6/1$E$-8\rightarrow 3$E$-5$ & $40{\rm dB}/20{\rm dB}$\\
2 & 1000 & 100  & $8$E$-4$ & $5.6$E$-4\rightarrow56$ & $\kappa_{\rm opt}$  & $40{\rm dB}$\\
3 & 1000 & $50\rightarrow1000$  & $8$E$-4$ & $10$ & $\kappa_{\rm opt}$  & $40{\rm dB}$\\
4 & 1000 & 100  & $4$E$-4$ & $10$ & $0.1\kappa_{\rm opt}\rightarrow10\kappa_{\rm opt}$ & $40{\rm dB}/20{\rm dB}$\\
5 & 1000 & 100  & $2$E$-4\rightarrow4$E$-4$ & $10$ & $\kappa_{\rm opt}$  & $40{\rm dB}$\\ \hline
\end{tabular}
\end{scriptsize}
\end{scriptsize}
\end{center}
\end{table*}

\begin{figure}
  \centering
  \includegraphics[width=4in]{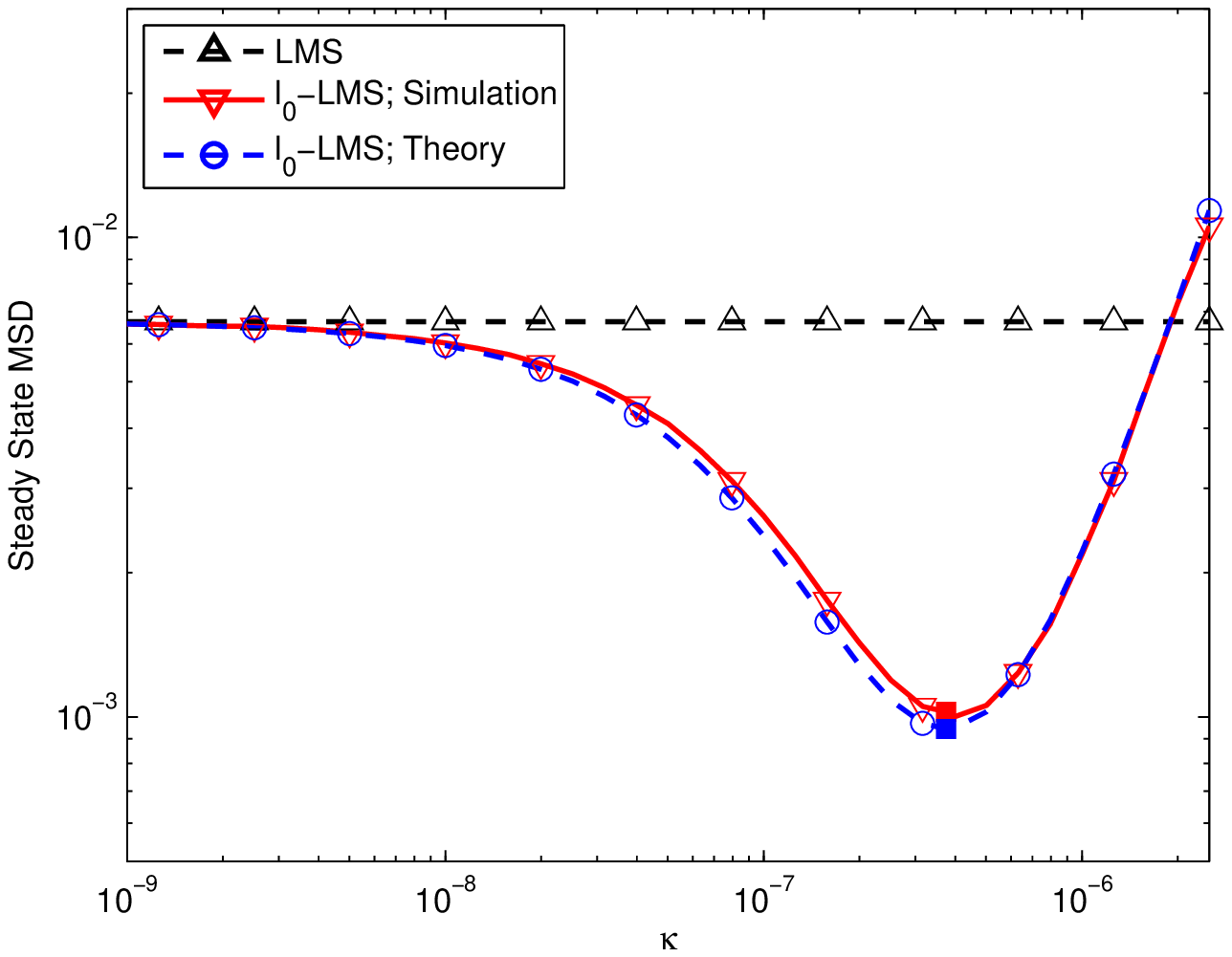}\\
  \caption{Steady-state MSD of LMS and $l_0$-LMS (with respect to different $\kappa$), {where SNR is $40$dB and the solid square denotes $\kappa_{\rm opt}$}.}\label{Fig:Steady_MSD_Sweep_Kappa_40dB}
\end{figure}

\begin{figure}
  \centering
  \includegraphics[width=4in]{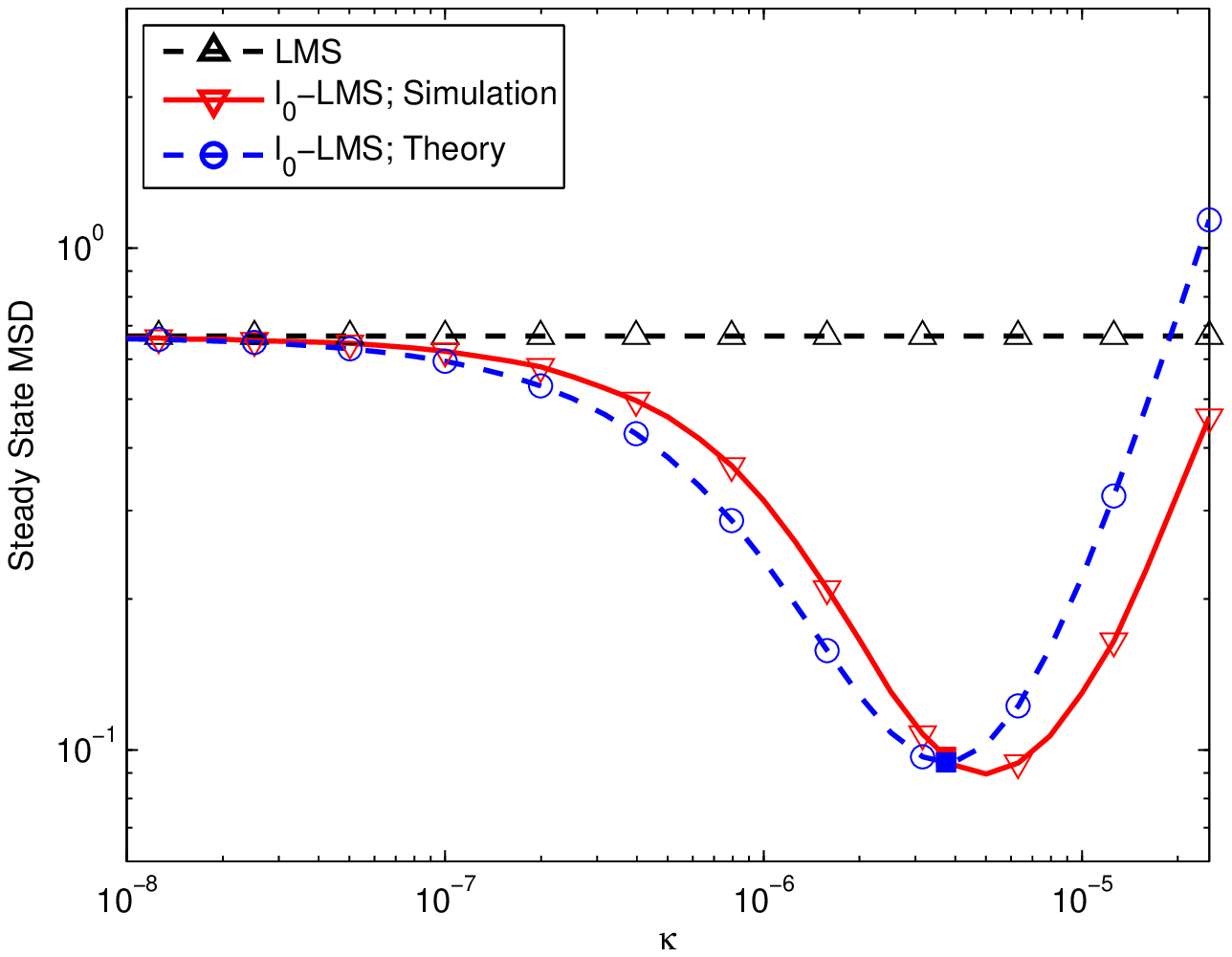}\\
  \caption{Steady-state MSD of LMS and $l_0$-LMS (with respect to different $\kappa$), {where SNR is $20$dB and the solid square denotes $\kappa_{\rm opt}$}.}\label{Fig:Steady_MSD_Sweep_Kappa_20dB}
\end{figure}

In the first experiment, the steady-state performance with respect to $\kappa$ is considered.
Referring to Fig.~\ref{Fig:Steady_MSD_Sweep_Kappa_40dB}, the theoretical steady-state MSD of $l_0$-LMS is in
good agreement with the experiment results when SNR is $40$dB. With the
growth of $\kappa$ from $10^{-9}$,
 the steady-state MSD decreases
at first, which means proper zero-point attraction is helpful for sufficiently reducing the
amplitude of tap-weights in $\mathcal{C}_0$. On the other hand, larger $\kappa$ results in more
intensity of zero-point attraction item and increases the bias of small coefficients
$\mathcal{C}_S$. Overlarge $\kappa$ causes too much bias, thus deteriorates the overall
performance. From (\ref{KappaBest}), $\kappa_{\rm {opt}}=3.75\times 10^{-7}$ produces the minimized
steady-state MSD, which is marked with a square in
Fig.~\ref{Fig:Steady_MSD_Sweep_Kappa_40dB}. Again, simulation result tallies with analytical
value well. When SNR is $20$dB, referring to Fig.~\ref{Fig:Steady_MSD_Sweep_Kappa_20dB}, the theoretical result also
predicts the trend of MSD well. However, since the assumptions (v) and (vi) do not hold well in
low SNR case, the theoretical result has {perceptible} deviation from the simulation result.

\begin{figure}
  \centering
    \includegraphics[width=4in]{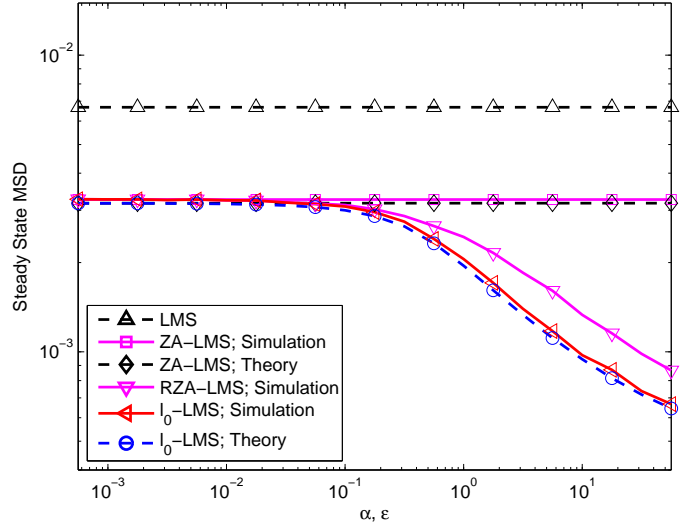}\\
    \caption{Steady-state MSD of LMS, ZA-LMS, RZA-LMS (with respect to different $\varepsilon$), and $l_0$-LMS (with respect to different $\alpha$), where $\varepsilon$ equals $\alpha${. Parameters} $\rho$ and $\kappa$ are chosen as optimal for RZA-LMS and $l_0$-LMS, respectively.}\label{Fig:Steady_MSD_Sweep_Alpha_40dB}
\end{figure}

In the second experiment, the effect of parameter $\alpha$ on steady-state performance is
investigated. Please refer to Fig.~\ref{Fig:Steady_MSD_Sweep_Alpha_40dB} for results.
RZA-LMS is also tested for performance comparison, with
its parameter $\rho$ chosen as optimal values which are obtained by experiments. For the sake of
simplicity, the parameter $\varepsilon$ in (\ref{RZAGradApprx}) is set the same as $\alpha$.
Simulation results confirm the validity of the theoretical analysis. With very small $\alpha$, all
tap-weights are attracted toward zero-point and the steady-state MSD is nearly independent. As
$\alpha$ increases, there are a number of taps fall in the attraction range while the others are
out of it. Consequently, the total bias reduces. Besides, the results for ZA-LMS are also
considered in this experiment, with the optimal parameter $\rho$ proposed in {\it Remark 5}. It is
shown that $l_0$-LMS always yields superior steady-state performance than ZA-LMS; moreover, in
scenario where $\alpha$ approaches $0$, the MSD of $l_0$-LMS  tends to that of ZA-LMS.
In the parameter range {of this experiment}, $l_0$-LMS shows better steady-state performance than RZA-LMS.

\begin{figure}
  \centering
  \includegraphics[width=4in]{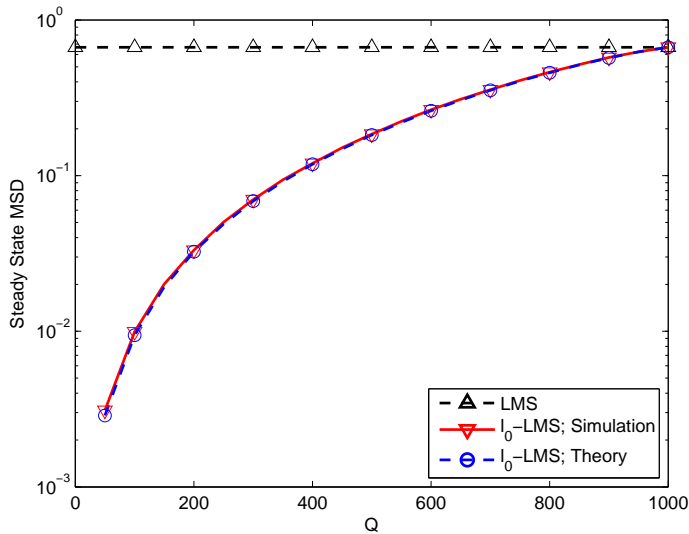}\\
  \caption{Steady-state MSD of LMS and $l_0$-LMS (with respect to different total non-zeros taps $Q$), where $\kappa$ is chosen as optimal.}\label{Fig:Steady_MSD_Sweep_Q}
\end{figure}

The third experiment studies the effect of non-zero coefficients
number on steady-state deviation. Please refer to Fig.
\ref{Fig:Steady_MSD_Sweep_Q}. It is readily accepted that $l_0$-LMS
with optimal $\kappa$ outperforms traditional LMS in steady state.
The fewer the non-zero unknown coefficients are, the more effectively
$l_0$-LMS draws tap-weights towards zero. Therefore, the
effectiveness of $l_0$-LMS  increases with the sparsity of the
unknown system. When $Q$ exactly equals $L$, its performance with
optimal $\kappa$ already attains that of standard LMS, indicating
that there is no room for performance enhancement of $l_0$-LMS for a
totally non-sparse system.

\begin{figure}
  \centering
  \includegraphics[width=4in]{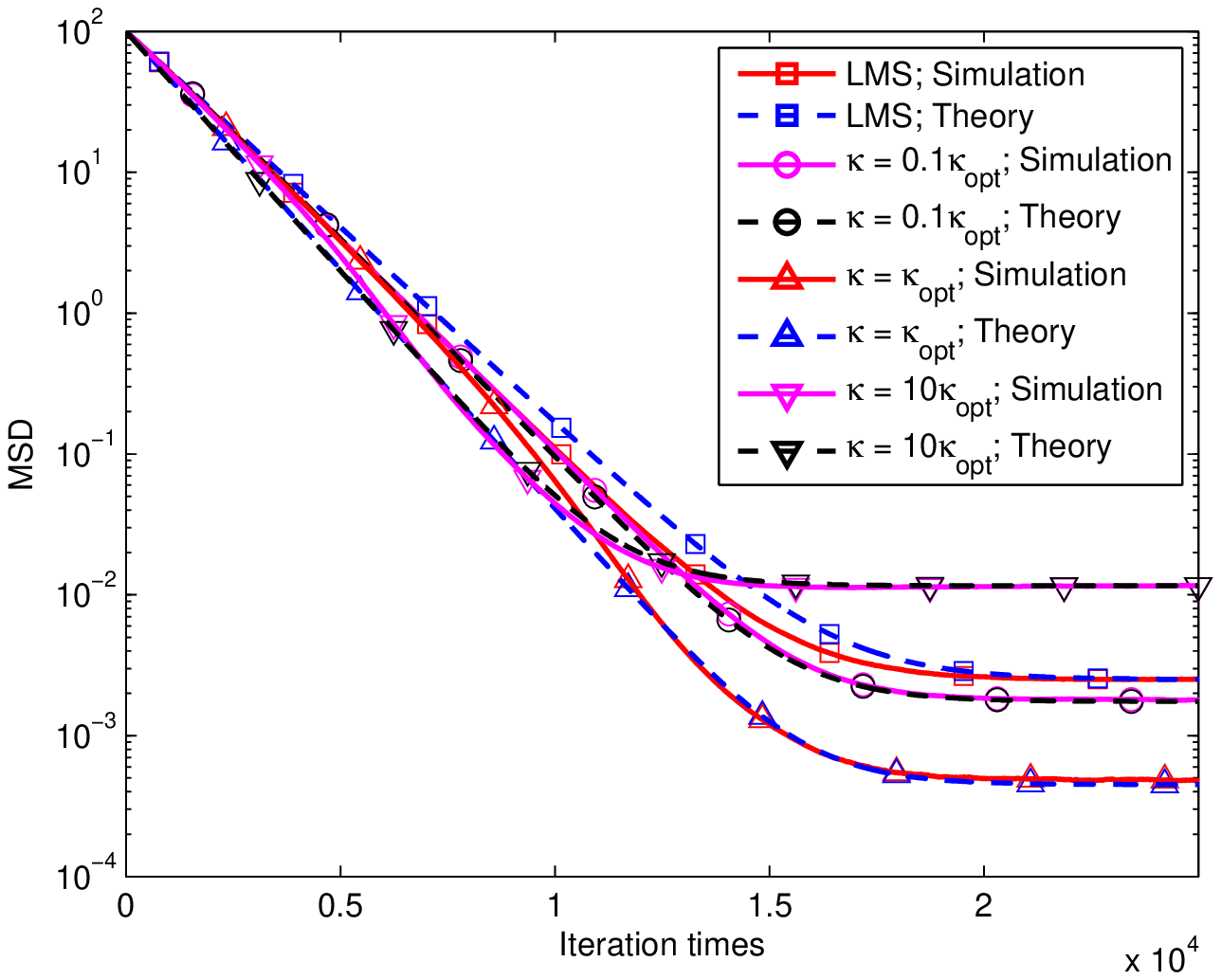}\\
  \caption{MSD convergence of LMS and $l_0$-LMS (with respect to different $\kappa$), where SNR is $40$dB.}
  \label{Fig:Converegence_MSD_Sweep_Kappa_40dB}
\end{figure}

\begin{figure}
  \centering
  \includegraphics[width=4in]{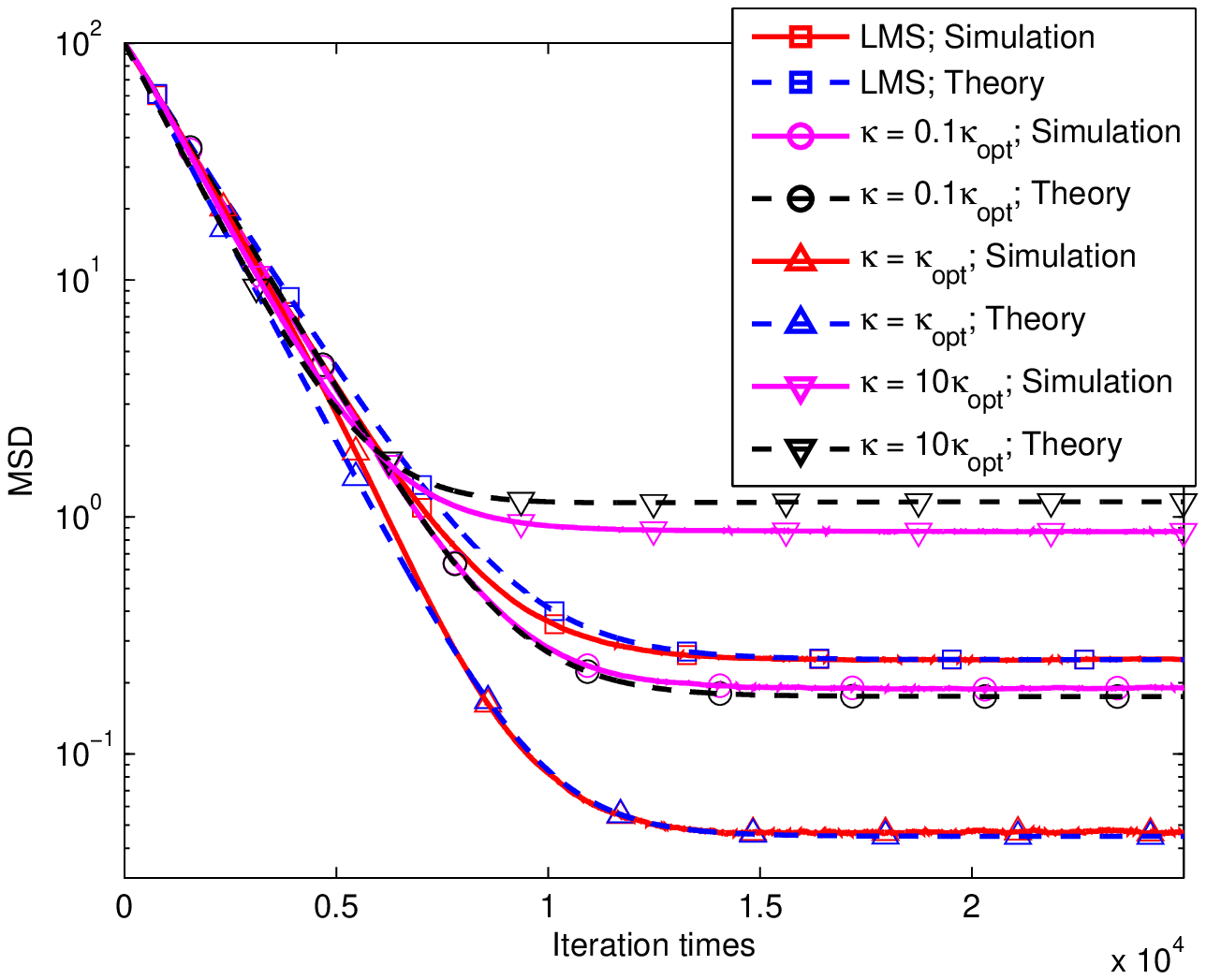}\\
  \caption{MSD convergence of LMS and $l_0$-LMS (with respect to different $\kappa$), where SNR is $20$dB.}
  \label{Fig:Converegence_MSD_Sweep_Kappa_20dB}
\end{figure}

The fourth experiment is designed to investigate convergence process with respect to $\kappa$.
Also, the learning curve of the standard LMS is simulated. When SNR is $40$dB, the results in
Fig. \ref{Fig:Converegence_MSD_Sweep_Kappa_40dB} demonstrate that our theoretical analysis of
convergence process is generally in good accordance with simulation. It can be observed that
different $\kappa$ results in differences in both steady-state MSD and the convergence rate. Due to
more intense zero-attraction force, larger $\kappa$ results in higher convergence rate; but too
large $\kappa$ can have bad steady-state performance for too much bias of small coefficients.
Moreover, $l_0$-LMS outperforms standard LMS in convergence rate for all parameters we run, and
also surpasses it in steady-state performance when $\kappa$ is not too large. When
SNR is 20dB, Fig.~\ref{Fig:Converegence_MSD_Sweep_Kappa_20dB} also shows similar trend about
how $\kappa$ influences the convergence process; however, since the low SNR scenario breaks
assumptions (v) and (vi), the theoretical results and experimental results differ to some extent.

\begin{figure}
  \centering
  \includegraphics[width=4in]{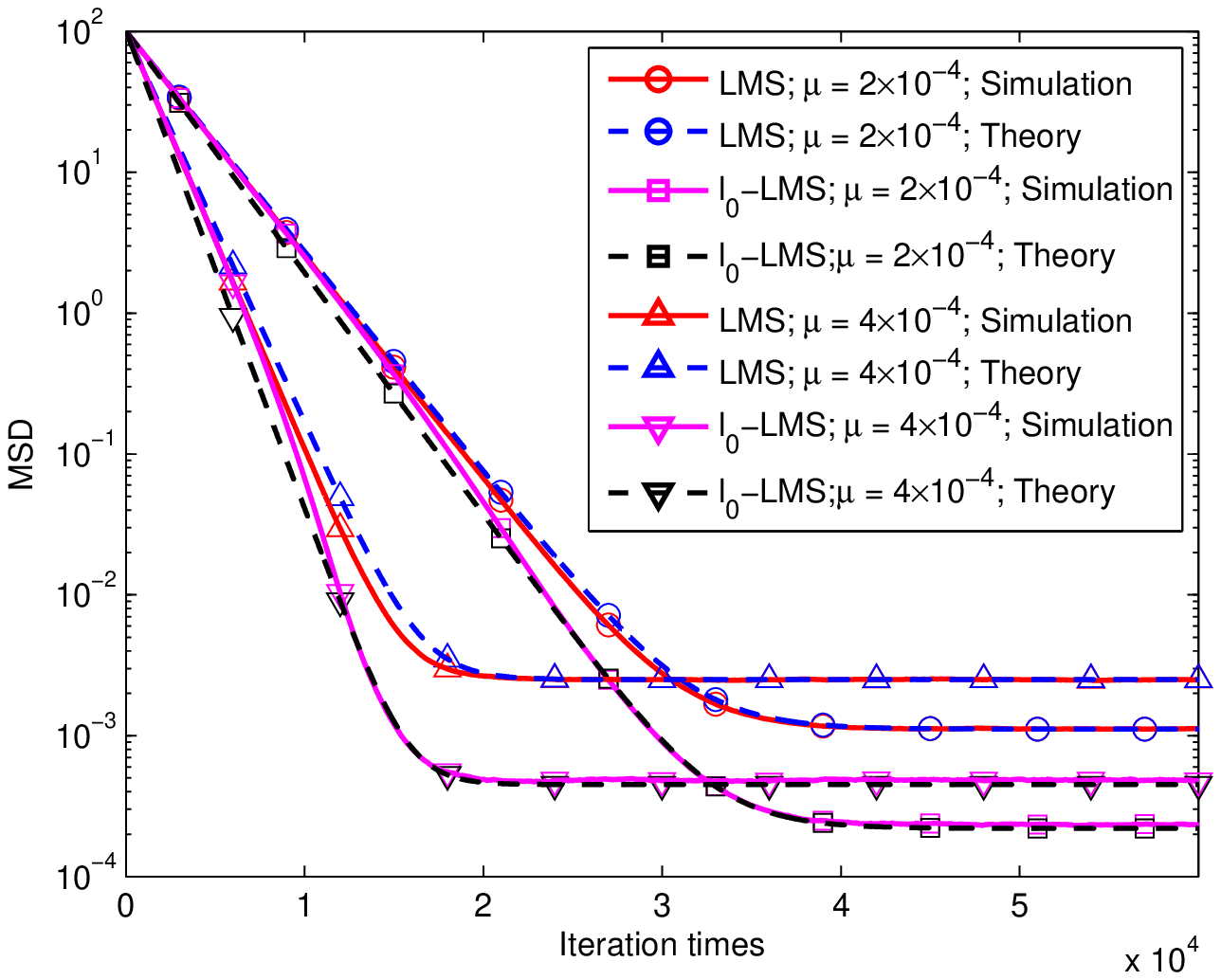}\\
  \caption{MSD convergence of LMS and $l_0$-LMS with respect to different step sizes $\mu$, where $\kappa$ is chosen as optimal for $l_0$-LMS.}\label{Fig:Converegence_MSD_Sweep_Mu}
\end{figure}

The fifth experiment demonstrates convergence process for various
step sizes, with the comparison of LMS and $l_0$-LMS. Please refer to
Fig. \ref{Fig:Converegence_MSD_Sweep_Mu}. Similar to traditional LMS,
smaller step size yields slower convergence rate and less
steady-state MSD. Therefore, the choice of step size should seek a
balance between convergence rate and steady-state performance.
Furthermore, the convergence rate of $l_0$-LMS is {faster} than
that of LMS when their step sizes are identical.

\section{Conclusion}\label{Section:Conclusion}
The {comprehensive} mean square performance analysis of $l_0$-LMS algorithm
is presented in this paper, including both steady-state and
convergence process. The adaptive filtering taps are firstly
classified into three categories based on the zero-point attraction
item, and then analyzed separately. With the help of {some assumptions which are reasonable in a wide range}, the steady-state MSD is
finally deduced and the convergence of instantaneous MSD is approximately
predicted. Moreover, a parameter selection rule is put forward to
minimize the steady-state MSD and theoretically it is shown that
$l_0$-LMS with optimal parameters is superior than traditional LMS
for sparse system identification. The all-round theoretical results are
verified in a large range of parameter setting through numerical simulations.

\appendix

\appendixtitleon
\begin{appendices}

\section{Expressions of constants}
\label{Const}

In order to make the main body simple and focused, the explicit
expressions of some constants used in derivations are
listed here.

All through this work, four constants of
\begin{align}
    \Delta_L &= 2-(L+2)\mu P_x\label{DefineDeltaL}\\
    \Delta_Q &= 2-(Q+2)\mu P_x\label{DefineDeltaQ}\\
    \Delta_0 &= 1-\mu P_x\label{DefineDelta0}\\
    \Delta'_0 &= 2-\mu P_x\label{DefineDelta01}
\end{align}
are used to simplify the expressions.

To evaluate the zero-point attracting strength, with respect to the sparsity of the unknown system coefficients, two kinds of strengthes are defined as
\begin{align}
G({\bf s}) &= \left<g({\bf s}), g({\bf s})\right> = \sum\limits_{k \in
\mathcal{C}_S}\!\!g^2(s_k),\label{AttractingStrength}\\
G'({\bf s}) &= \left<{\bf s}, g({\bf s})\right> = \sum\limits_{k \in
\mathcal{C}_S}\!\!s_kg(s_k),\label{AttractingStrength1}
\end{align}
which are utilized everywhere in this work. Considering the
attraction range, it can be readily accepted that these strengthes
are only related to the small coefficients, other than the large
ones and the zeros.

In Lemma \ref{lemma}, ${\bf
A}=\left\{a_{ij}\right\}$ is defined as
\begin{equation}
    \left[\!\!\begin{array}{cc}1-\mu P_x\Delta_L & -\sqrt{\frac{8}{\pi}}\frac{\alpha\kappa}{\omega}\Delta_0 \\ (L-Q)\mu^2P_x^2 &
    1\!-\!2\mu P_x\Delta_0\!-\!\sqrt{\frac{8}{\pi}}\frac{\alpha\kappa}{\omega}\Delta_0 \end{array}\!\!\right],\label{Matrix_A}
\end{equation}
and
\begin{equation}
    {{\bf b}_n}=[b_{0,n}, b_{1,n}]^{\rm T},\label{Vector_b}
\end{equation}
where
\begin{align*}
b_{0,n} =& L\mu^2 P_x P_v
+(L-Q)\left(4\alpha^2\kappa^2-\sqrt{8/ \pi}\alpha\kappa\omega\Delta_0\right)\nonumber\\
&+\frac{\kappa^2(\Delta'_0-2\Delta_0^{n+1})}{\mu
P_x}G({\bf s})-2\kappa\Delta_0^{n+1}G'({\bf s}), \\
b_{1,n} =& (L-Q)\left(\mu^2 P_x P_v+ 4\alpha^2\kappa^2-\sqrt{8/
\pi}\alpha\kappa\omega\Delta_0\right),
\end{align*}
where $\omega$ is the solution to (\ref{Quadratic_EQU}).

In Theorem \ref{Theo2}, the constants $\lambda_3$ and $c_3$  are
\begin{align}
\lambda_3 =& \Delta_0,\label{lambda3}\\
c_3 =& -\frac{2\kappa\Delta_0 \left(\mu P_x-2\mu^2
P_x^2+\sqrt{\frac{8}{\pi}}\frac{\alpha\kappa}{\omega}\Delta_0\right)}{\mu P_x \det\left( \lambda_3 {\mathbf{I}}
-{\mathbf{A}}\right)}\left(\kappa G({\bf s}) + \mu P_x G'({\bf s})\right). \label{C3}
\end{align}

In Corollary \ref{Coro:Opt_Kappa}, the constants $\{\beta_i\}$ are
\begin{align}
\beta_0 =& \mu P_x \Delta'_0\Delta_LG({\bf s})
+4\alpha^2\Delta_Q\left(\mu P_x \Delta_L+\frac{\Delta_0 \Delta_Q}{\pi}\right)
,\label{beta0}\\
\beta_1 =& \frac{\Delta'_0G({\bf s})
+ 4(L-Q)\alpha^2\left(\mu P_x +\frac{2\Delta_0\Delta_Q}{\pi\Delta_L}\right)}{\mu^2 P_x^2\Delta_L},\label{beta1}\\
\beta_2 =& \frac{4\alpha (L-Q)}{\mu^2 P_x^2\Delta_L^2}\sqrt{\frac{\Delta_0\beta_0}{\pi}} \label{beta2},\\
\beta_3 =& 2\mu^3 P_x^2 P_v  \Delta_0 \Delta_L /
\beta_0.\label{beta3}
\end{align}

In Appendix \ref{Effect_Sparsity} and \ref{Effect_Mu}, the constants
$\{\eta_i\}$ are
\begin{align}
&\eta_0 = \frac{16P_v \alpha^2\Delta_0^2}{\pi \mu P_x^2 \Delta_L^3},\label{eta0}&
&\eta_1 = \frac1{\mu^2 P_x^2 \Delta_L},\\
&\eta_2 = \frac{(L-Q)\beta_0}{\Delta_L\Delta_Q} ,&
&\eta_3 = \frac{4\alpha^2(L-Q)\Delta_0\Delta_Q}{\pi\Delta_L},\\
&\eta_4 = \frac{G({\bf s})\Delta'_0\Delta_L}{\Delta_Q},& \\
&\eta_5 = 4\alpha^2 P_x\mu+2G({\bf s})L,&
&\eta_6 = \frac{16\alpha^2
L}{\pi \Delta_L}.\label{eta6}
\end{align}

\section{Proof of Theorem \ref{Theo1}}
\label{Proof_Steady_MSD}
\begin{proof}
Denote $D_n$ to be MSD at iteration $n$, and ${\bf{R}}_n$ to be the second moment matrix of ${\bf
h}_n$, respectively,
\begin{align}
      D_n=&\overline{{\bf h}^{\rm{T}}_n{\bf h}_n},\label{MSD}\\
{\bf R}_n=&\overline{{\bf h}_n {\bf h}^{\rm{T}}_n}.\label{SecondMoment}
\end{align}
Substituting (\ref{h_Update}) into (\ref{SecondMoment}), and expanding the term
$\overline{{\bf x}_n {\bf x}^{\rm{T}}_n{\bf h}_n {\bf h}^{\rm{T}}_n{\bf x}_n {\bf x}^{\rm{T}}_n}$
into three second moments using the Gaussian moment factoring theorem
\cite{ref:Analysis_ComplexNarrowBand}, one knows
\begin{align}
{\bf R}_{n+1}=&(1-2\mu P_x\Delta_0){\bf R}_n + \mu^2 P_x^2\cdot {\rm {tr}}\left\{{\bf R}_n\right\} {\bf I} + \mu^2 P_x P_v {\bf I} \nonumber\\
            &+ \kappa\Delta_0\overline{{\bf h}_n g({\bf w}^{\rm{T}}_n)
            } +\kappa\Delta_0\overline{ g({\bf w}_n){\bf h}_n^{\rm{T}}}
            + \kappa^2 \overline{g({\bf w}_n) g({\bf w}^{\rm{T}}_n)  }.\label{R_Update}
\end{align}
Using
the fact that $D_n={\rm{tr}}\left\{ {\bf R}_n \right\}$, one has
\begin{align*}
D_{n+1}=(1-\mu P_x\Delta_L)D_n + L \mu^2 P_x P_v+2\kappa\Delta_0\overline{{\bf h}^{\rm{T}}_n g({\bf w}_n) }
      + \kappa^2 \overline{ \|g({\bf w}_n)\|_2^2  }.
\end{align*}
Consequently, the condition needed to ensure convergence is
$\left|1-\mu P_x\Delta_L\right|<1$
and (\ref{ConvCondition}) is derived
directly, which is the same as standard LMS and similar with the conclusion
in \cite{ref:ADF_CS}.

Next the steady-state MSD will be derived. Using (\ref{R_Update})
and considering the $k$th diagonal element, one knows
\begin{align}
&\overline{h^2_{k,\infty}} = \frac{\mu^2\!P_x^2\!D_\infty\!\!+\!\!\mu^2\!P_x\!P_v\!+\!2\kappa\Delta_0\overline{h_{k,\infty} g(w_{k,\infty}) }\!\!+\!\!\kappa^2\overline{g^2(w_{k,\infty})}}{2\mu P_x\Delta_0} .\label{MSD_Comp_SteadyState}
\end{align}
To develop $\overline{h^2_{k,\infty}}$, one should first investigate
two items, namely $\overline{h_{k,\infty} g(w_{k,\infty})}$ and $\overline{g^2(w_{k,\infty})}$ in
(\ref{MSD_Comp_SteadyState}). For $k\in \mathcal{C}_L$, from assumption (vi)
one knows $|w_{k,\infty}|>{1}/{\alpha}$, thus
\begin{equation}
\overline{h_{k,\infty} g(w_{k,\infty})
}=\overline{g^2(w_{k,\infty})}=0.\label{EhgEgg_LC}
\end{equation}
For small coefficients, considering assumptions (v) and (vi), formula (\ref{l_0GradApprx})
implies {$g(w_{k,\infty})$ is a locally linear function with slope $2\alpha^2$, which results in}
\begin{equation}
g(w_{k,\infty})=g(s_{k})+2\alpha^2 h_{k,\infty}.\nonumber
\end{equation}
Thus, it can be shown
\begin{align}
\overline{h_{k,\infty}g(w_{k,\infty})}&=2 \alpha^2 \overline{h^2_{k,\infty}} + g(s_k) \overline{h_{k,\infty}},\label{Ehg_SC}\\
\overline{g^2(w_{k,\infty})}&=4 \alpha^4 \overline{h^2_{k,\infty}}+
g^2(s_k) + 4\alpha^2 g(s_k)\overline{h_{k,\infty}},\label{Egg_SC}
\end{align}
where $\overline{h_{k,\infty}}$ is derived in (\ref{h_SteadyState_Detail}).

Then turning to $k\in \mathcal{C}_0$, it is readily known that $h_{k,\infty}=w_{k,\infty}$ in
this case. Thus, from assumptions (iv) and (vi), the following results can be derived from the
property of Gaussian distribution,
\begin{align}
\overline{h_{k,\infty}g(w_{k,\infty})}&=2 \alpha^2 \overline{h^2_{k,\infty}}-2\alpha
\overline{|h_{k,\infty}|}\nonumber\\
&=2 \alpha^2 \overline{h^2_{k,\infty}}
 - 4\alpha \sqrt{\overline{h^2_{k,\infty}}}/\sqrt{2\pi}, \label{Ehg_ZC}\\
\overline{g^2(w_{k,\infty})}&=4 \alpha^4 \overline{h^2_{k,\infty}} -
8\alpha^3\overline{|h_{k,\infty}|}+4\alpha^2\nonumber\\&=
4 \alpha^4 \overline{h^2_{k,\infty}} - 16\alpha^3
\sqrt{\overline{h^2_{k,\infty}}}/\sqrt{2\pi}+4\alpha^2.\label{Egg_ZC}
\end{align}
Combining assumption (iii), (\ref{h_SteadyState_Detail}) and
(\ref{EhgEgg_LC})$\sim$(\ref{Egg_SC}), one can know the equivalency
between (\ref{MSD_Comp_SteadyState}) and following equations for $k$
in $\mathcal{C}_L$, $\mathcal{C}_S$, and $\mathcal{C}_0$,
respectively,
\begin{align}
2\mu P_x\Delta_0\overline{h^2_{k,\infty}}
 -
\mu^2 P_x^2D_\infty - \mu^2 P_x P_v &= 0, \quad k\in
\mathcal{C}_L,\label{MSD_Comp_SteadyState_LC}\\
2\mu P_x\Delta_0\overline{h^2_{k,\infty}}
  - \mu^2
P_x^2D_\infty - \mu^2 P_x P_v
-\kappa^2 g^2(s_k) \left(2/\mu/P_x -1\right) &= 0,\quad k\in
\mathcal{C}_S,\label{MSD_Comp_SteadyState_SC}\\
2\mu
P_x\Delta_0\omega^2
+8\alpha\kappa\Delta_0\omega/\sqrt{2\pi}
- \mu^2 P_x^2D_\infty - \mu^2 P_x P_v - 4\alpha^2\kappa^2 &= 0,\quad
k\in \mathcal{C}_0,\label{MSD_Comp_SteadyState_ZC}
\end{align}
where $\omega$ denotes $\sqrt{\overline{h^2_{k,\infty}}}, k\in \mathcal{C}_0$ for simplicity.
Summing (\ref{MSD_Comp_SteadyState_LC}) and (\ref{MSD_Comp_SteadyState_SC}) for all $k\in
\mathcal{C}_L\bigcup \mathcal{C}_S$, and noticing that
$$
\sum\limits_{k\in \mathcal{C}_L\bigcup
\mathcal{C}_S}\overline{h^2_{k,\infty}}=D_\infty-(L-Q)\omega^2,
$$
it could be derived that
\begin{equation}
    D_\infty=\frac{2(L-Q)\Delta_0}{\Delta_Q} \omega^2+ \frac{Q\mu P_v}{\Delta_Q}
       +\frac{\kappa^2 \Delta'_0}{\mu^2 P_x^2 \Delta_Q}G({\bf s}),\label{MSD_Expression_Old}
\end{equation}
where $G({\bf s})$ is introduced in (\ref{AttractingStrength}). Combining
(\ref{MSD_Expression_Old}) and (\ref{MSD_Comp_SteadyState_ZC}), it can be reached that $\omega$ is
defined by the following equation
\begin{align}
2\mu P_x\Delta_0\Delta_L\omega^2
+\frac{8 \alpha \kappa \Delta_0\Delta_Q}{\sqrt{2\pi}}\omega-2\mu^2 P_x P_v \Delta_0
-4\alpha^2 \kappa^2\Delta_Q - \kappa^2 \Delta'_0G({\bf
s})&=0.\label{Quadratic_EQU}
\end{align}
Finally, (\ref{MSD_Expression}) is achieved after solving the
quadratic equation above and a series of formula transformation on
(\ref{MSD_Expression_Old}). Thus, the proof of Theorem \ref{Theo1}
is completed.
\end{proof}

\section{Proof of Corollary \ref{Coro:Opt_Kappa}}
\label{Best_Parameters}
\begin{proof}
By defining $\theta = \arctan\left(\kappa/\sqrt{\beta_3}\right)$,
(\ref{MSD_Expression}) becomes
\begin{equation}
D_\infty = \mu P_v L/\Delta_L-\beta_1 \beta_3
+\beta_3\cdot f(\sin(\theta))/2, \label{MSD_Theta}
\end{equation}
where $f(x)$ is defined as
$$
f(x)= \frac{\beta_1-\beta_2}{1-x}+\frac{\beta_1+\beta_2}{1+x},\quad
x\in (0,1).
$$
Next we want to find $x_{\rm {opt}}\in (0,1)$ which minimizes
$f(x)$. Forcing the derivative of $f(x)$ with respect to $x$ to be
zero, it can be obtained that
$$
0< x_{\rm {opt}}
=\frac{\sqrt{\beta_1+\beta_2}-\sqrt{\beta_1-\beta_2}}{\sqrt{\beta_1+\beta_2}+\sqrt{\beta_1-\beta_2}}<1.
$$
Combining $\theta_{\rm {opt}} = \arcsin (x_{\rm {opt}})$ and
substituting $\theta_{\rm {opt}}$ in (\ref{MSD_Theta}), corollary
\ref{Coro:Opt_Kappa} can be finally achieved.
\end{proof}

{
\section{Proof of Corollary \ref{Coro:Steady_Q}}\label{Effect_Sparsity}
\begin{proof}
From (\ref{MSDLMS}), (\ref{MSD_Opt}), (\ref{beta1}), (\ref{beta2}),
and (\ref{eta0}), it can be obtained that
\begin{equation}
D^{\rm min}_\infty = D^{\rm LMS}_\infty-
\frac{\eta_0}{\frac{\beta_1}{(L-Q)^2}+\frac{\sqrt{\beta_1^2-\beta_2^2}}{(L-Q)^2}}.\label{MSD_Opt_eta}
\end{equation}
Note neither the $D^{\rm LMS}_\infty$ defined in (\ref{MSDLMS}) nor
$\eta_0$ defined in (\ref{eta0}) is dependent on $Q$ or $G({\bf
s})$, thus the focus of the proof is the denominator in
(\ref{MSD_Opt_eta}). In the following, we will analyze the two items
in the denominator separately and obtain their monotonicity.
The first item in the denominator is
\begin{align}
&\frac{\beta_1}{(L-Q)^2} = \frac{1}{\mu^2 P_x^2 \Delta_L} \Bigg(\frac{\Delta'_0}{(L-Q)^2}G({\bf s})+\frac{4\alpha^2}{L-Q} \left(\mu P_x +
\frac{2\Delta_0}{\pi}\right) +\frac{8\alpha^2\mu P_x\Delta_0}{\pi
\Delta_L}\Bigg),\label{beta1_Coro:Steady_Q}
\end{align}
From (\ref{beta1_Coro:Steady_Q}), it has already shown that
$\beta_1/(L-Q)^2$ is increasing with respect to $Q$ and $G({\bf
s})$.
Next we consider the second item. It can be obtained beforehand that
$\beta_1$ and $\beta_2$ equal
$\eta_1\left(\eta_2+\eta_3+\eta_4\right)$ and $2\eta_1\sqrt{\eta_2
\eta_3}$, respectively. Thus, one has
\begin{equation}
\frac{\sqrt{\beta_1^2-\beta_2^2}}{(L-Q)^2} =
\eta_1\sqrt{\frac{\eta_4^2+2\eta_4(\eta_2+\eta_3)+(\eta_2-\eta_3)^2}{(L-Q)^4}}.\label{beta1beta2_Coro:Steady_Q}
\end{equation}
Further notice that
\begin{align*}
\eta_2-\eta_3=& \mu P_x(L-Q)\left(4\alpha^2+\frac{\Delta'_0
}{\Delta_Q}G({\bf s})\right),\\
\eta_2+\eta_3=&4(L-Q)\alpha^2\left(\mu P_x +\frac{2\Delta_0\Delta_Q}{\pi \Delta_L}\right) 
+ \frac{\mu P_x(L-Q)\Delta'_0}{\Delta_Q}G({\bf s}),
\end{align*}
it can be proved that all of the three items in the square root of
(\ref{beta1beta2_Coro:Steady_Q}) are increasing with respect to $Q$
and $G({\bf s})$; thus the second item in the denominator is
monotonic increasing with respect to $Q$ and $G({\bf s})$.
Till now, the monotonicity of
$D^{\rm min}_\infty$ with respect to $Q$ and $G({\bf s})$ has been
proved.
Last, in the special scenario where $Q$ exactly equals $L$, it
can be obtained that $D^{\rm min}_\infty$ is identical to $D^{\rm
LMS}_\infty$; thus $D^{\rm
LMS}_\infty$ is larger than the minimum steady-state MSD of the
scenario where $Q$ is less than $L$. In sum, Corollary
\ref{Coro:Steady_Q} is proved.
\end{proof}
\section{Proof of Corollary \ref{Coro:Steady_mu}}\label{Effect_Mu}
\begin{proof}
For a sparse system in accordance with (\ref{Sparse_Condition}),
$\{\eta_i\}$ defined in Appendix \ref{Const} are approximated by
\begin{align*}
&\eta_0 \approx \frac{16P_v \alpha^2}{\pi \mu P_x^2 \Delta_L^3}, &
&\eta_2 \approx \frac{8\alpha^2 L}{\pi\Delta_L} + L\mu P_xG({\bf s})+4\alpha^2\mu P_x L,\\
&\eta_3 \approx \frac{8\alpha^2 L}{\pi\Delta_L}, & &\eta_4 \approx
\Delta_LG({\bf s}).
\end{align*}
Substituting $\{\eta_i\}$ in $\{\beta_i\}$ of (\ref{MSD_Opt_eta}),
with the approximated expressions above,
(\ref{MSD_Simple_Sparse_SmallMu}) is finally derived after
calculation.
Next we show $D^{\rm min}_\infty$ in
(\ref{MSD_Simple_Sparse_SmallMu}) is monotonic increasing with
respect to $\mu$. Since (\ref{MSD_Simple_Sparse_SmallMu}) is
equivalent with
\begin{equation}
D^{\rm min}_\infty\!\!\approx\!\!
 \frac{ P_v
L\left(\eta_5+\sqrt{\eta_5^2+\frac{32\alpha^2 L}{\pi}G({\bf
s})}\right)}{\frac{16\alpha^2 L}{\pi\mu}
\!+\!\Delta_L\!\!\left(\!\!\frac{\eta_5}{\mu}\!+\!\!\sqrt{\left(\frac{\eta_5}{\mu}\right)^2\!\!\!+\!\!\frac{32\alpha^2
L}{\pi\mu^2}G({\bf s})}\right)}.\label{MSD_Simple_Appdendix}
\end{equation}
it can be directly observed from (\ref{eta6}) that larger $\mu$
results in larger numerator as well as smaller denominator in
(\ref{MSD_Simple_Appdendix}), which both contribute to the fact that
$D^{\rm min}_\infty$ is monotonic increasing with respect to $\mu$.
Thus, the proof of Corollary \ref{Coro:Steady_mu} is arrived.
\end{proof}
}

\section{Relationship with ZA-LMS}
\label{Relationship_Sparse_Standard_LMS}
 When $2\alpha\kappa=\rho$
remains a constant while $\alpha$ approaches zero, from
  (\ref{w_Update}), (\ref{l_0GradApprx}), and (\ref{l_1GradApprx}), it is obvious that
the recursion of $l_0$-LMS becomes that of ZA-LMS. Furthermore, one
can see that $g^2(x)$ equals $4\alpha^2+o(\alpha^2)$. From the
definition, it can be shown $\mathcal{C}_L$ is an empty set when
$\alpha$ approaching zero. Consequently,
\begin{equation}
G({\bf s})=|\mathcal{C}_S|\cdot
(4\alpha^2+o(\alpha^2))=4\alpha^2Q+o(\alpha^2). \label{gSC_sum_L1}
\end{equation}
Combine (\ref{MSD_Expression_Old}), (\ref{Quadratic_EQU}), and
(\ref{gSC_sum_L1}), then after quite a series of calculation, the
explicit expression of steady-state MSD becomes
\begin{align}
D^{\rm ZA}_\infty=&-\frac{(L-Q)\rho\sqrt{\Gamma}}{\sqrt{2\pi}\mu^2 P_x^2 \Delta_L^2}+\frac{2\rho^2(L-Q)\Delta_0\Delta_Q}{\pi\mu^2 P_x^2 \Delta_L^2}
+\frac{\rho^2\left(\mu LP_x + 2Q\Delta_0\right)+L\mu^3 P_x^2
P_v}{\mu^2 P_x^2 \Delta_L},\label{MSD_L1_Expression}
\end{align}
where $\Gamma$ is the discriminant of quadratic equation
(\ref{Quadratic_EQU}),
\begin{equation}
\Gamma =8\rho^2\Delta_Q^2\Delta_0^2/\pi +16\mu P_x \Delta_L
\Delta_0^2\left(\rho^2(Q+1)+\mu^2 P_x P_v\right).\label{Gamma_ZA}
\end{equation}
Through a series of calculation, it can be proved that
(\ref{MSD_L1_Expression}) is equivalent with
(\ref{Shi_Sparse_LMS_MSD}) obtained
in \cite{ref:ShiKun}. Thus, the steady-state MSD in ZA-LMS could be
regarded as a particular case of that in $l_0$-LMS.

\section{Proof of Lemma \ref{lemma}}
\label{Proof_Instantaneous_MSD}
\begin{proof}
 From (\ref{R_Update}), the update formula is
\begin{align}
\overline{h^2_{k,n+1}}=&(1-2\mu P_x\Delta_0)\overline{h^2_{k,n}} + \mu^2 P_x^2D_n+ \mu^2 P_x P_v+2\kappa\Delta_0\overline{h_{k,n} g(w_{k,n}) } + \kappa^2 \overline{g^2(w_{k,n})}.\label{Coeff_Update}
\end{align}

Since LMS algorithm has fast convergence rate, it is reasonable to
suppose most filter tap-weights will get close to the corresponding
system coefficient very quickly; thus, the classification of
coefficients $s_k$ could help in the derivation of the convergence
situation of $h_{k,n}$.

For $k\in \mathcal{C}_L$, from
assumption (vi),
(\ref{Coeff_Update}) takes the form
\begin{equation}
\overline{h^2_{k,n+1}}=(1-2\mu P_x\Delta_0)\overline{h^2_{k,n}}
+ \mu^2 P_x^2D_n + \mu^2 P_x P_v , k\in \mathcal{C}_L
.\label{LC_Conv}
\end{equation}

For $k\in \mathcal{C}_S$ the mean convergence is firstly derived and
then the mean square convergence is deduced.  Take expectation in
(\ref{h_Update}),
and combine assumptions (iii), (v),
and (vi), one knows
$$
\overline{h_{k,n+1}}=\Delta_0\overline{h_{k,n}}+\kappa
g(s_k),k\in \mathcal{C}_S.
$$
Since $h_k(0) = -s_k$, one can finally get
\begin{equation}
\label{Mean_Conv_SC} \overline{h_{k,n}} = \frac{\kappa g(s_k)}{\mu
P_x }-\frac{\mu P_x s_k +\kappa g(s_k)}{\mu P_x}\Delta_0^n, k\in \mathcal{C}_S.
\end{equation}
Combining (\ref{Coeff_Update}) , (\ref{Mean_Conv_SC}) and employing
assumption (iii), it can be achieved
\begin{align}\
\overline{h^2_{k,n+1}}=&(1-2\mu P_x\Delta_0 )\overline{h^2_{k,n}}+ \mu^2 P_x^2 D_n + \mu^2 P_x P_v  + 2\kappa\Delta_0 g(s_k) \overline{h_{k,n}} +\kappa^2 g^2(s_k) , k\in \mathcal{C}_S.\label{SC_Conv}
\end{align}

Next turn to $k\in \mathcal{C}_0$. From assumption (iv), the following formula can be attained
employing the steady state result and first-order Taylor expansion
$$
\overline{\left| h_{k,n} \right|}=\sqrt{2\overline{h^2_{k,n}}/\pi}
\approx \left(\overline{h^2_{k,n}}/\omega + \omega\right)/\sqrt{2\pi}, k\in \mathcal{C}_0,
$$
where $\omega = \sqrt{\overline{h^2_{k,\infty}}}, k\in
\mathcal{C}_0$, which is the solution
to equation (\ref{Quadratic_EQU}).
Finally,
with assumption (iii) we have
\begin{align}\
\overline{h^2_{k,n+1}}=&\left(1-2\mu P_x\Delta_0 - \sqrt{\frac{8}{\pi}}\frac{\alpha\kappa}{\omega} \Delta_0\right)\overline{h^2_{k,n}} + \mu^2 P_x^2 D_n  \nonumber\\
    & + \mu^2 P_x P_v +4\alpha^2\kappa^2 - \sqrt{\frac{8}{\pi}}\alpha\kappa\omega\Delta_0 ,k\in \mathcal{C}_0 .\label{ZC_Conv}
\end{align}

Considering $\Omega_n=\sum_{k\in \mathcal{C}_0}\overline{h^2_{k,n}}$, and combine (\ref{LC_Conv}),(\ref{Mean_Conv_SC}),(\ref{SC_Conv}), and
(\ref{ZC_Conv}), one can obtain (\ref{D_Omega_Iter}) after a series of derivation. As for the initial value, since $\mathbf{w}_0=0$, by
definition we have $D_0=\|{\bf s}\|_2^2$ and $\Omega_0 = 0$. Thus, Lemma \ref{lemma} is reached.
\end{proof}

\section{Proof of Theorem \ref{Theo2}}
\label{lemma_To_Theo2}
\begin{proof}
The vector $\mathbf{b}_n$ in (\ref{Vector_b}) could be denoted as
\begin{equation}
{\mathbf{b}}_n = \left[\!\!
\begin{array}{c}
\hat{b}_{00}+\widehat{b}_{01}\lambda_3^{n}\\
\widehat{b}_{1}\end{array} \!\!\right] ,  \label{b_simp}
\end{equation}
where $\lambda_3$ is defined in (\ref{lambda3}) and
$\widehat{b}_{00},\widehat{b}_{01},\widehat{b}_1$ are constants.
Take $z$-Transform for (\ref{D_Omega_Iter}), it can be derived that
$$
\left[\!\!\!
\begin{array}{c}
D(z)\\
\Omega(z)
\end{array}\!\!\!
\right] =(z\mathbf{I}-\mathbf{A})^{-1} z \left[\!\!\!
\begin{array}{c}
D_0\\
\Omega_0\end{array}\!\!\!\right]
+(z\mathbf{I}-\mathbf{A})^{-1}{\mathbf{b}(z)},
$$
where $z>1$. Then combine the definition of $\{\lambda_i\}$ in
Theorem \ref{Theo2} and the above
results, it is further derived
$$
D(z) = \sum_{i=0}^{3}\frac{c_i}{1- \lambda_iz^{-1}},
$$
where $\lambda_0=1$ and $\{c_i\}$ are constants. Take the inverse $z$-Transform and notice the definition of $D_\infty$, it finally yields
$$
D_n = D_\infty+c_1\lambda_1^{n}+c_2\lambda_2^{n}+c_3\lambda_3^{n}.
$$
Thus we have completed the proof of (\ref{D_Conv_ClosedForm}). By forcing the equivalence between (\ref{D_Conv_ClosedForm}) and Lemma
\ref{lemma}, the expression of $c_3$ could be solved as (\ref{C3}).
\end{proof}

\section{Proof of Corollary \ref{Coro:Fast_Conv_mu}}
\label{Proof_Coro:Fast_Conv_mu}
\begin{proof}
Define function
$$
p(x)=\det\left|x{\mathbf{I}}-{\mathbf{A}} \right|,\ x\in\mathbb{R},
$$
then the roots of $p(x)$ are eigenvalues of matrix $\mathbf{A}$.
From (\ref{Matrix_A}), it can be shown
$$
\det \left|a_{00} {\mathbf{I}}-{\mathbf{A}} \right| = \det
\left|a_{11} {\mathbf{I}}-{\mathbf{A}} \right| =
-a_{01}a_{10}>0,
$$
and
$$
\det \left|\frac{a_{00}+a_{11}}{2} {\mathbf{I}}-{\mathbf{A}}
\right|
\le-\frac{1}{4}\left(L\mu^2
P_x^2-\sqrt{\frac{8}{\pi}}\frac{\alpha\kappa}{\omega}\Delta_0\right)^2\le0,
$$
where $\{a_{ij}\}$ denote the entries of $\bf A$. Thus, we know $p(a_{11})>0$ and $p\left(\frac{a_{00}+a_{11}}{2}\right)\le0$, which indicates that one root of quadratic function $p(x)$ is within the interval $\left(a_{11}, \frac{a_{00}+a_{11}}{2}\right]$. Similarly, another root of $p(x)$ is in $\left[\frac{a_{00}+a_{11}}{2},a_{00}\right)$. Thus, it can be concluded
that the eigenvalues of $\mathbf{A}$ are both in $\mathbb{R}$ and
satisfy
\begin{equation}
a_{11}<\lambda_1\le \frac{a_{00}+a_{11}}{2}\le \lambda_2<a_{00}= 1-\mu P_x\Delta_L.
\label{lambda12_Bound_Full}
\end{equation}
For large step size scenario of $1< \mu (L+2)P_x <2$, (\ref{lambda3}) and (\ref{lambda12_Bound_Full}) yield
$$
\max\left\{\lambda_1,\lambda_2,\lambda_3\right\}<1-\mu P_x\Delta_L.
$$

Through comparison between (\ref{D_Conv_ClosedForm}) and (\ref{D_Conv_ClosedForm_LMS}), one can know for large $\mu$, all the three transient
items in MSD convergence of $l_0$-LMS has faster attenuation rate than LMS, leading to acceleration of convergence rate.
\end{proof}

\end{appendices}

\section*{Acknowledgement}

The authors wish to thank Laming Chen and four anonymous reviewers for their helpful comments to
improve the quality of this paper.

\end{document}